\documentclass[aps,pra,twocolumn,floatfix,superscriptaddress]{revtex4-2}

\usepackage{amsmath}
\usepackage{amssymb}

\usepackage{subfigure}
\usepackage{graphicx}

\begin{document}

\title{Proximity effect and spatial Kibble-Zurek mechanism in atomic Fermi gases with inhomogeneous pairing interactions}

\author{Bishal Parajuli}
\affiliation{Department of Physics, University of California, Merced, CA 95343, USA.}

\author{Chih-Chun Chien}
\email{cchien5@ucmerced.edu}
\affiliation{Department of Physics, University of California, Merced, CA 95343, USA.}

\begin{abstract}
Introducing spatially tunable interactions to atomic Fermi gases makes it feasible to study two phenomena, the proximity effect and spatial Kibble-Zurek mechanism (KZM), in a unified platform. While the proximity effect of a superconductor adjacent to a normal metal corresponds to a step-function quench of the pairing interaction in real space, the spatial KZM is based on a linear drop of the interaction that can be modeled as a spatial quench.
After formulating the Fermi gases with spatially varying pairing interactions by the Bogoliubov-de Gennes equation, we obtain the profiles of the pair wavefunction and its correlation function to study their penetration into the noninteracting region. For the step-function quench,
both correlation lengths from the pair wavefunction and its correlation function follow the BCS coherence length and exhibit the same scaling behavior.
In contrast, the scaling behavior of the two correlation lengths are different in the spatial quench, which then allows more refined analyses of the correlation lengths from different physical quantities. Moreover, adding a weakly interacting bosonic background does not change the scaling behavior. We also discuss relevant experimental techniques that may realize and verify the inhomogeneous phenomena.
\end{abstract}

\maketitle

\section{Introduction}
Cold atoms have been a versatile platform for studying fundamental physics and simulating complex many-body physics \cite{pitaevskii2003bose,Pethick-BEC,Ueda-book}. 
Developments of spatially-resolved manipulations of the interactions between atoms beyond conventional means \cite{exp-optical-feshbach1,exp-optical-feshbach2,inhomo-condensate1,inhomo-condensate2,inhomo-condensate3,inhomo-condensate4,PhysRevLett.115.155301,PhysRevLett.125.183602} allow cold-atoms to exhibit interesting inhomogeneous phenomena. 
Here we study atomic Fermi superfluid with controllable inhomogeneous interactions to revisit two seemingly different phenomena, the proximity effect of superconductors~\cite{degennes-proximity}  and spatial Kibble-Zurek mechanism \cite{Dziarmaga-rev}, in an integrated framework. The origin of both phenomena comes from the concept of quantum phase transition, where driving a parameters of the Hamiltonian across a critical point causes a fundamental change of the ground state~\cite{sachdev_2011}. The pairing interaction will be the parameter separating the broken-symmetry Fermi superfluid
and the symmetric normal gas in this study.

When a superconductor (SC) is in contact with a normal metal (NM), the Cooper pairs from the SC penetrate into the NM with a characteristic length determined by the BCS coherence length \cite{Falk-PE,degennes-proximity}, a phenomenon known as the proximity effect. The NM acquires some properties of the SC, such as a reduction in the resistance and the ability to carry a supercurrent \cite{clarke-proximity}.  
The proximity effect results from a sudden change of the pairing interaction across the SC-NM interface, so it may be thought of as a phase transition in space. The proximity effect in other heterostructures have been extensively studied, including a superconductor-quasicrystal hybrid ring \cite{PhysRevB.100.165121}, disordered and quasi-periodic systems \cite{PE-GRai}, superconducting thin films \cite{PE-RLkobes} and normal metal- superconducting slab \cite{PE-Ssharma}. Experimental \cite{PhysRevB.73.094507,PhysRevB.81.094503} and theoretical \cite{PhysRevB.91.165142,PhysRevB.94.104511,PE-Ssharma,PE-GCsire} studies of niobium-gold layers suggest that the proximity effect may create topological superconductivity.  
In addition, experimental data of granular SC-NM structures are shown to agree with the theory \cite{degennes-proximity} in the weak coupling limit. 
Furthermore, there have been extensive research on the proximity effect of ferromagnet-superconductor heterostructures \cite{SC-FM-rev}, which may give rise to the Majorana bound state  \cite{PhysRevLett.100.096407}.

On the other hand, the Kibble-Zurek mechanism (KZM) ~\cite{kibble-1976,kibble-1980,Zurek-1985,Zurek-1996} studies the reaction of a system crossing a continuous phase transition. The systems can be driven by a time-dependent or time-independent ramp. 
The KZM has inspired a plethora of theoretical \cite{Zurek-1997,Zurek-1998,Zurek-1999,Zurek-2002,Dziarmaga-doiqpt,QPT-Dziarmaga-2005,PhysRevD.81.025017,Uhlmann_2010,Dziarmaga-2022,Dziarmaga-rev,QIM-Jaschke-2017,adiabtic-dynamics-2005,Damski-2005,QD-Legget-2005,QPT-Zurek-2005,Zurekptis,PhysRevA.97.033626,PhysRevA.75.023603,PhysRevB.86.144521,PhysRevB.95.104306,PhysRevA.103.013310,PhysRevLett.116.155301,Damski_2009} and experimental \cite{Monaco-2002,Ulm-2013,Pyka-2013,ALGaunt-2015,QPT-Brawn-2015,QPT-mott-2011,QKZM-Nature-2019,QKZM-prl-2016,TFIM-BWLi-2022,KZM-colloidal-monolayer-2015,KZM-FermiSF-2019,KZM-vortices-2021} studies to verify or compare the KZ scaling. The majority of the investigations have focused on time-dependent quenches of the parameters, where the excitations follow a power-law dependence of the transition rate \cite{QPT-Dziarmaga-2005,QPT-Zurek-2005,QIM-Jaschke-2017,Dziarmaga-2022}, including  the Bose-Hubbard model \cite{PhysRevA.75.023603,PhysRevB.86.144521,PhysRevB.95.104306,PhysRevA.97.033626,PhysRevA.103.013310} and spinor Bose-Einstein condensates (BEC) \cite{PhysRevLett.99.120407,PhysRevLett.116.155301,Damski_2009}. As a system approaches a critical point within $\epsilon$, the reaction time $\tau$ diverges as $\tau\sim |\epsilon|^{-\nu z}$, which determines how fast the system can react.  
After the system is driven into the broken-symmetry phase, the density of topological excitaitons reflects the frozen correlation length $\xi\sim\tau_Q^{\frac{\nu}{1+\nu z}}$, where $\tau_Q$ is the characteristic quench time, and $\nu$ and $z$ are the critical exponents from the corresponding phase transition. Dynamics of the ground state of Fermi superfluid following a time quench has also been studied~\cite{Chien10}. Additionally, there have been studies beyond the mean-number analysis of the KZM~\cite{PhysRevLett.122.080604,PhysRevResearch.2.033369,PhysRevLett.124.240602,Commun.Phys.3.44}.

Meanwhile, the time-independent KZM, also known as the spatial KZM, considers a linear ramp of the interaction and analyzes the scaling in the vicinity of a critical point in real space separating a broken-symmetry phase and a symmetric phase.  
The spatial KZM has been formulated and summarized in Refs.~\cite{Zurekptis,Dziarmaga-doiqpt,Damski_2009} with applications to the quantum transverse-field Ising model~ \cite{Zurekptis,Dziarmaga-doiqpt} and spin-1 BEC~ \cite{Damski_2009}.
In the spatial KZM,
the order parameter or its correlation function penetrates into the symmetric phase. Different from the abrupt drop of the interaction in the proximity effect, the linear ramp of the interaction introduces additional length scale. For a typical continuous phase transition in a uniform system, the correlation length $\xi$ diverges near the critical point as $\xi\sim|\epsilon|^{-\nu}$~\cite{ChaikinP.M1995PoCM}. In the spatial KZM, the correlation length freezes out within the transition region where the interaction is linearly ramped to zero, which in turn determines the characteristic length of the penetration into the symmetric phase. The correlation length on the symmetric phase side follows the scaling behavior $\xi\sim \alpha^{-\nu/(1+\nu)}$. Here $\alpha$ measures the slope of the ramp in real space, which is the counterpart of the quench rate in a time-dependent quench. 
Importantly, 
the spatial KZM keeps the whole system in equilibrium, which is very different from the nonequilibrium nature of the time-dependent KZM. The trade-off is that the spatial KZM only determines one exponent $\nu$ instead of two in the time-dependent KZM. Previous theoretical studies analyzed possible structures of atomic Fermi gases with inhomogeneous pairing interactions~\cite{Chien12,PhysRevB.98.144508} but did not explore the spatial KZM.

By formulating the Bogoliubov-de Gennes (BdG) equation \cite{bogoliubov1947theory,BdG-book} for two-component atomic Fermi gases with inhomogeneous pairing interactions dropping from a constant value to zero, we extract the BCS coherence length, pair-wavefunction correlation length, and pair-pair correlation length and compare their scaling behavior. If the pairing interaction drops to zero abruptly, we call it a step-function quench, and the system simulates the proximity effect in SC-NM heterostructures. If the interaction drops according to a linear ramp, we call it a spatial quench and show the system exhibits the spatial KZM. While the BCS coherence length dominates in the step-function results, the correlation lengths from the pair wavefunction and its correlation function lead to different scaling behavior in the spatial quench. The scaling of the correlation length from the pair wavefunction follows the spatial KZM based on the BCS theory at $T=0$, but that of the pair correlation function exhibit observable deviation. Therefore, the spatial KZM of Fermi superfluid is able to differentiate different correlation lengths.
By adding a bosonic background in the miscible phase, we confirm the scaling of the coherence and correlation lengths of the fermions stay intact. Importantly, we show the scaling behavior can be established in the ground states of finite systems in equilibrium, which extends available probes of quantum phase transitions. 

The rest of the paper is organized as follows. Sec.~\ref{sec:theory} briefly reviews the mean-field theory of two-component Fermi gases with attractive interactions and its applications to ultracold fermionic atoms. Sec.~\ref{sec:quench} describes the two quench protocols and their relevance to previous studies. Sec.~\ref{sec:result} presents the correlation lengths and their scaling behavior in the two quench protocols and explains the mechanism behind the observations. Sec.~\ref{sec:implication} contrasts the subtle differences between the spatial KZM in Fermi gases and magnetic systems and discuss possible experimental techniques for realizing and measuring the inhomogeneous phenomena studied here. Finally, Sec.~\ref{sec:conclusion} concludes our work.

\section{Theoretical background} \label{sec:theory}

\subsection{Two-component fermions with attractive interaction}
The second-quantization Hamiltonian for two-component  fermions labeled by $\sigma=\uparrow,\downarrow$ with an effective attractive interaction $V_{eff}$ is given by
\begin{align}\label{eq::Hamiltonian}
    \mathcal{H}&=\sum_{\sigma}\int dr \psi_{\sigma}^{\dagger}(r)h_{\sigma}(r)\psi_{\sigma}(r) \nonumber \\
    &-\frac{1}{2}\sum_{\sigma,\sigma'} \iint dr dr'V_{eff}(r,r')\psi_{\sigma}^{\dagger}(r)\psi_{\sigma'}^{\dagger}(r')\psi_{\sigma'}(r')\psi_{\sigma}(r).
\end{align}
Here $\psi_{\sigma}^{\dagger}(r)$ (or $\psi_{\sigma}(r)$) is the fermion creation (or annihilation) operator with spin $\sigma$ at location $r$, and $h_\sigma(r)=-\frac{\hbar^2}{2m}\nabla^2+V_{ext}(r)-\mu_\sigma$.
The BCS mean-field approximation then leads to \cite{Fetter_book}
\begin{align}\label{Eq:HBCS}
     \mathcal{H}_{BCS}&=\sum_{\sigma}\int dr \psi_{\sigma}^{\dagger}(r)h_{\sigma}(r)\psi_{\sigma}(r) \nonumber \\
     &+\iint dr dr'(\Delta(r,r') \psi_{\uparrow}^{\dagger}(r)\psi_{\downarrow}^{\dagger}(r')+h.c) \nonumber \\
     &+\iint dr dr'|\Delta(r,r')|^2/V_{eff}(r,r').
\end{align}
The gap function $\Delta(r,r')$ is defined as
\begin{align}\label{gap-original}
    \Delta(r,r')&=-V_{eff}(r-r')\langle\psi_{\downarrow}(r')\psi_{\uparrow}(r)\rangle.
\end{align}
Here $\langle O\rangle$ is the ensemble average of operator $O$. We will focus on the case where the two components have equal population with the same chemical potential $\mu$.

\subsection{Bogoliubov-de Gennes equation}
The Bogoliubov-de Gennes transformation is given by \cite{bogoliubov1947theory,BdG-book},
\begin{eqnarray}
     \psi_{\uparrow}(r)&= \sum_{\tilde{n}}[u_\uparrow^{\tilde{n}1}(r)\gamma_{\tilde{n}1}-v_\uparrow^{\tilde{n}2*}(r)\gamma_{\tilde{n}2}^{\dagger}], \nonumber \\
\psi_{\downarrow}(r)&= \sum_{\tilde{n}}[u_\downarrow^{\tilde{n}2}(r)\gamma_{\tilde{n}2}+v_\downarrow^{\tilde{n}1*}(r)\gamma_{\tilde{n}1}^{\dagger}],
\end{eqnarray}
which diagonalizes the BCS Hamiltonian~\eqref{Eq:HBCS} into the form
\begin{equation}\label{Eq:HBdG}
\mathcal{H}_{BCS}=\sum_{\tilde{n}w} E_{\tilde{n}w} \gamma_{\tilde{n}w}^\dagger \gamma_{\tilde{n}w} +E_g,
\end{equation}
where, $w=1,2$ represents the two-component of the quasi-particle operators. $E_g$ is the ground state energy given by $E_g=\frac{|\Delta|^2}{V_{eff}}+\sum_{\tilde n,w}(\epsilon_{\tilde nw}-E_{\tilde{n}w})$. Here $\epsilon_{\tilde {n}w}$ is the non-interacting ($V_{eff}=0$) counterpart of the excitation energy $E_{\tilde {n}w}$. 
The coefficients of $\gamma_{\tilde{n}w}$ and $\gamma_{\tilde{n}w}^\dagger$ in the canonical transformation can be determined by the Bogoliubov-de Gennes (BdG) equation \cite{degennes-sc}.
In the absence of spin-orbit coupling,  
the BdG equation is block-diagonalized into two sets of equations. Explicitly,
\begin{equation}\label{eq:subset1}
    \sum_{r'} \begin{pmatrix}
    h_{\uparrow}(r,r')&\Delta(r,r')\\
    \Delta^*(r',r)&-h^*_{\downarrow}(r,r')
    \end{pmatrix}
    \begin{pmatrix}
    u^{\tilde{n}1}_{\uparrow}(r')\\
    v^{\tilde{n}1}_{\downarrow}(r')
    \end{pmatrix}
    =E_{\tilde{n}1}\begin{pmatrix}
    u^{\tilde{n}1}_{\uparrow}(r)\\
    v^{\tilde{n}1}_{\downarrow}(r)
    \end{pmatrix},
\end{equation}
and a similar matrix equation for $\tilde{n}2$. Here $h_{\sigma}(r,r')=h_{\sigma}(r)\delta(r-r')$.

The wavefunction $u_{\uparrow}$ is coupled only to the wave function $v_\downarrow$ and similarly for $u_\downarrow$ and $v_\uparrow$.
A symmetry of the two sets of the BdG equations in the absence of spin-orbit coupling leads to
$\begin{pmatrix}
    u^{\tilde{n}2}_{\downarrow}(r)\\
    v^{\tilde{n}2}_{\uparrow}(r)
    \end{pmatrix}
    =\begin{pmatrix}
    v^{\tilde{n}1 *}_{\downarrow}(r)\\
    -u^{\tilde{n}1}_{\uparrow}(r)
    \end{pmatrix}$
and $E_{\tilde{n}2}=-E_{\tilde{n}1}$. The symmetry implies that
we can solve one of the two sets of equations and focus on the positive-energy states.
The quasi-particle operators obey $\langle\gamma_{\tilde{n}w}^\dagger\gamma_{\tilde{m}v}\rangle=\delta_{\tilde{n}\tilde{m}}\delta_{wv}f(E_{\tilde{n}w})$ and $ \langle\gamma_{\tilde{n}w}\gamma_{\tilde{m}v}\rangle=\langle\gamma_{\tilde{n}w}^\dagger\gamma_{\tilde{m}v}^\dagger\rangle=0$. Here    $f(E_{\tilde{n}w})=[e^{E_{\tilde{n}w}/K_BT}+1]^{-1}$
is the Fermi distribution function. 
From now on, we drop the indices $1,2$ and $\uparrow,\downarrow$ from the quasi-particle wavefunctions.
The gap function (\ref{gap-original}) then becomes
$\Delta(r,r')=V_{eff}(r-r'){\sum_{\tilde{n}}}' u^{\tilde{n}}_{\uparrow }(r)v^{\tilde{n}*}_{\downarrow}(r')\tanh(E_{\tilde{n}}/k_B T)$.
Here ${\sum_{\tilde{n}}}'$ means the summation is over the positive-energy states.

\subsection{Atomic Fermi gases}
When applying the BCS theory to two component fermionic atoms, the two-body scattering length $a_{3D}$ serves as an indicator of the interaction between atoms~\cite{Pethick-BEC,Ueda-book}, which can be tuned by a magnetic field. For many-body systems, the effective interaction may be approximated by a contact interaction with coupling constant $g_{3D}$. Away from resonance, 
$g_{3D}=\frac{4\pi\hbar^2a_{3D}}{m}$.
However, Feshbach resonance has been used for studying BCS superfluids of cold atoms and the BCS-Bose-Einstein condensation (BEC) crossover~\cite{Pethick-BEC,Ueda-book}. Near a resonance, the renormalized interaction is
$    \frac{1}{g_{3D}} = \frac{m}{4 \pi \hbar^2 a_{3D}}-\frac{1}{V}\sum_k\frac{1}{2\epsilon_k}$.
Here $\epsilon_k$ is the dispersion of noninteracting fermions.
For fermionic superfluids of cold atoms, $a_{3D}<0$ indicates the conventional BCS superfluid while $a_{3D}>0$ indicates a condensate of tightly-bound pairs.

While the physics of proximity effect and spatial KZM is essentially 1D, we consider quasi-1D systems here for two reasons. Firstly, the Mermin-Wagner theorem~\cite{ChaikinP.M1995PoCM} rules out continuous long-range order in 1D, so Fermi superfluid in a quasi-1D setup is more appropriate. Secondly, the discussions here will be relevant to the elongated cigar-shape atomic clouds in experiments. In quasi-1D Fermi gases, the 1D effective coupling constant maybe expressed as \cite{Olshani} 
$g_{1D}=\frac{2\hbar^2a_{3D}}{m a_{\perp}^2}\frac{1}{1-A a_{3D}/a_{\perp}}$,
where $A$ is a constant associated with the confinement induced resonance and $a_{\perp}$ is the characteristic length in the transverse direction. The effective interactions 
 switch from attractive to repulsive at the confinement induced resonance $A=a_\perp/a_{3D}$. Therefore, $g_{1D}$ may be expressed as
\begin{equation}
    g_{1D}=-\frac{2\hbar^2}{ma_{1D}}
\end{equation} with the 1D scattering length given by
$    a_{1D}=-\frac{a_{\perp}^2}{a_{3D}}(1-Aa_{3D}/a_{\perp})$.
We remark that a quasi-1D BCS-BEC crossover occurs when the chemical potential changes sign because $a_{1D}$ is always positive. Hereafter we will drop the subscript $1D$ and implicitly assume strong confinements in the transverse directions.

The effective interaction in atomic Fermi gases is dominated by the contact interaction valid at low temperatures, so $V_{eff}(r-r')=-g(r)\delta(r-r')$. 
Thus, $\Delta(r,r')=\Delta(r',r)=\Delta(r)\delta(r-r')$. 
We consider equal population of the two components,  $N_\uparrow=N/2=N_\downarrow$, so $\mu_{\sigma}=\mu$.
For a two-component Fermi gas in a 1D box of length $L$ in the $x$ direction, we discretize the space as $x/L=[0,1]$ using $n_x$ grid points. $x_j=j \delta x$, where $\delta x=L/n_x$ and $j=0,1,2,....,n_x$. The Laplacian operator is represented by using the finite-difference method. In the discretized form, the BdG equation becomes 
\begin{equation}\label{discrete-BdG}
   \sum_j \begin{pmatrix}
        h_{ij}&\Delta_{ij}\\
        \Delta^*_{ij}& -h_{ij}
    \end{pmatrix}
    \begin{pmatrix}
        u_j^{\tilde{n}}\\
        v_j^{\tilde{n}}
    \end{pmatrix}
    =E_{\tilde{n}}\begin{pmatrix}
        u_i^{\tilde{n}}\\
        v_i^{\tilde{n}}
    \end{pmatrix}.
\end{equation}
Note that for s-wave pairing, $\Delta_{ij}=0$ if $i\neq j$. The BdG Hamiltonian has the size of $2n_x\times 2n_x$ and we only take the positive energy eigenstates for the calculations of the gap function and density.

The fermion density of each component is    $\rho_\sigma(x)=\langle\psi_\sigma^\dagger(x)\psi_\sigma(x)\rangle$, and the total density $\rho(x)=\sum_\sigma \rho_{\sigma}(x)$ 
becomes
\begin{equation}\label{Eq:rhoBdG}
\rho(x)=2{\sum_{\tilde n }}' |v_{\tilde n}(x)|^2.
\end{equation}
The total fermion number is $N=N_{\uparrow}+N_{\downarrow}=\int_0^L \rho(x)dx$.
The gap function is given by
\begin{equation}\label{Eq:DeltaBdG}
    \Delta(x)
    =-g(x){\sum_{\tilde n}}' u_{\tilde n}(x)v_{\tilde n}(x).
\end{equation}
However, we distinguish the pairing correlations from the gap function, which is necessary in studying Fermi gases with inhomogeneous interactions.
The pair wavefunction is \cite{Leggett}
\begin{equation}\label{Eq:F}
    F(x)=\langle\psi_\downarrow(x) \psi_\uparrow(x)\rangle={\sum_{\tilde n}}' u_{\tilde n}(x)v_{\tilde n}(x).
\end{equation}
We also consider the pair-pair correlation function given by
\begin{equation}\label{eq:correlation}
    C(r)=\overline{F(x)F(x+r)}.
\end{equation}
Here the over-line denotes an average over $x$. The correlation function is important in defining the critical exponent in homogeneous systems \cite{ChaikinP.M1995PoCM} and extracting the exponents in systems with inhomogeneous interactions.

\begin{figure}[t]
    \centering
    \includegraphics[width=0.9\columnwidth,keepaspectratio]{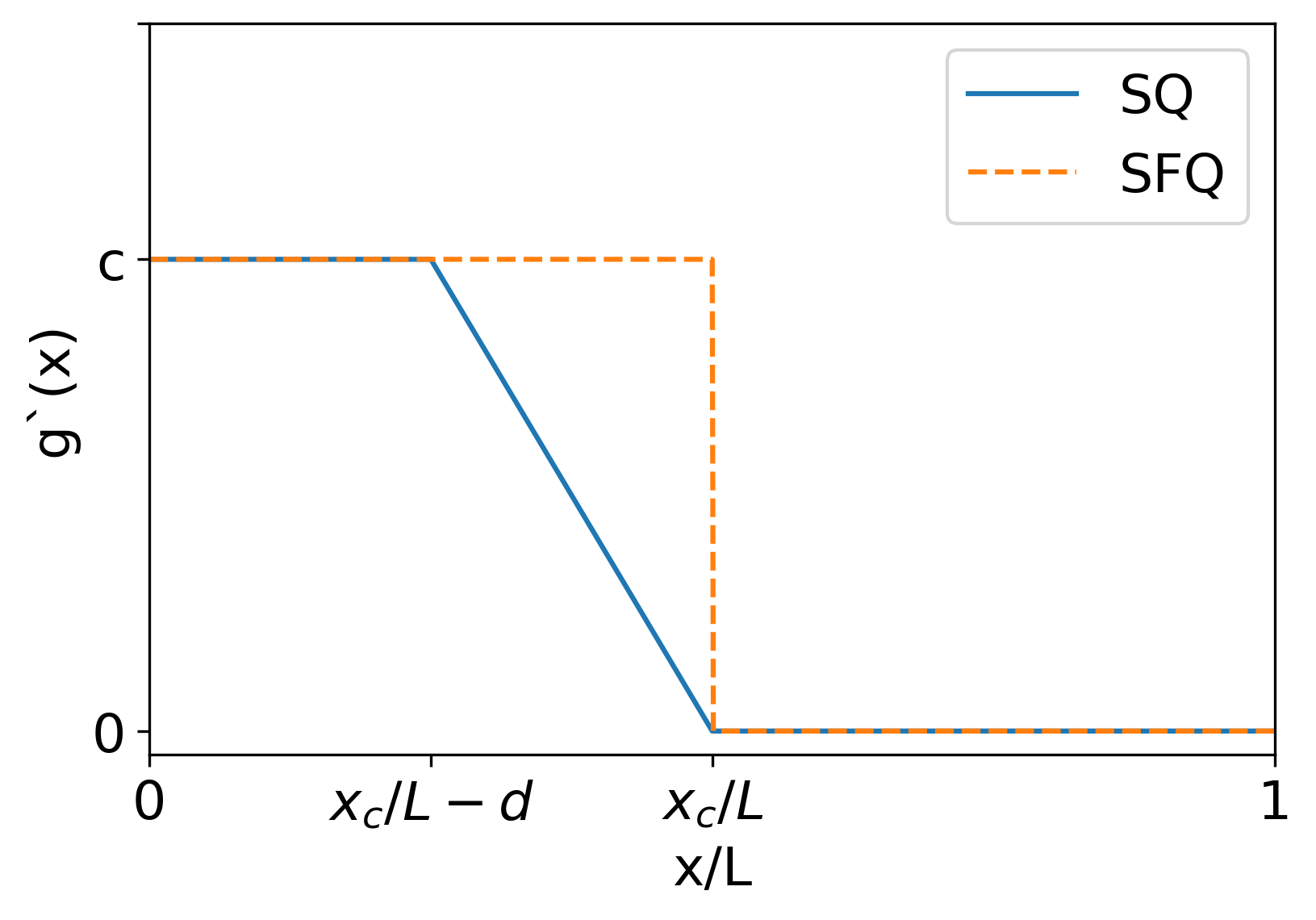}    \caption{Illustrations of the spatial quench (SQ, solid line) and step-function quench (SFQ, dashed line) of the pairing  interaction. }
    \label{fig:gff}
\end{figure}

\section{Interaction quench in real space}\label{sec:quench}
To study the analogues of the proximity effect and spatial KZM using atomic Fermi gases in a quasi 1D box potential of length $L$, we consider spatially dependent attractive interaction $g(x)$ between the two components. We use the Fermi energy $E_f^0$ and Fermi wavevector $k_f^0$ of a noninteracting Fermi gas with the same particle number to rewrite physical quantities in dimensionless forms. For example, the dimensionless interaction strength $g'(x)$ is defined by
$g(x)=-g'(x)E_f^0/k_f^0$.

\subsection{Step-function quench and proximity effect}
To simulate the sudden drop in the interaction, we consider the step-function quench of the pairing interaction that vanishes suddenly at $x=x_c$. For the step-function quench,
\begin{equation}
    g'(x)= 
\begin{cases}
    c, &   0 \leq x/L < x_c/L, \\
   
    0,              &  1>x/L>x_c/L.
\end{cases}
\end{equation}
We typically set $x_c=L/2$, where the order parameter vanishes. The interaction profile is illustrated in Fig.~\ref{fig:gff}.

In the study of proximity effect in a SC-NM junction, the pairing interaction is assumed to vanish across the interface. 
Previous studies \cite{clarke-proximity,degennes-proximity,Falk-PE} modeled the leakage of Cooper pairs from the superconductor into the normal metal with a characteristic length associated with the BCS coherence length. The decay of  $F(x)$ in the normal region at finite temperatures has the exponential form \cite{degennes-proximity,Falk-PE}
\begin{equation}\label{eq:exp-decay}
    F(x)\sim F_0 e^{-(x-x_c)/\xi_F}, ~T>0.
\end{equation}
Here $\xi_F$ is the correlation length associated with $F$. However, at zero temperature, $F(x)$ is no longer decaying exponentially with the distance  $y=x-x_c$ from the interface. Instead, it follows a power law $1/y$, as shown in Refs. \cite{degennes-proximity,Falk-PE,PE-Silvert}. Thus, the scaling behavior is
\begin{equation}\label{eq:power-law}
    \frac{F(x)}{\tilde{\Delta}}\frac{1}{k_f}\sim \frac{\xi_F}{x-x_c},~T=0.
\end{equation}
Here $\tilde{\Delta}=\Delta/E_f^0$ is the dimensionless bulk gap in the superfluid region.
The scaling behavior was obtained by solving the Gor'kov equation in Refs.~\cite{degennes-proximity,Falk-PE} and verified in SC-NM hybrid rings~\cite{PE-GRai}, superconducting thin films~\cite{PE-RLkobes}, niobium-gold layers~\cite{PE-GCsire}, and normal metal on top of a superconducting slab~\cite{PE-Ssharma}. The reason for the slower power-law decay of $F(x)$ into the normal metal at zero temperature is because thermal excitations are absent in restricting the penetration of Cooper pairs. We mention that Ref.~\cite{PhysRevB.98.144508} studies the proximity effect in atomic Fermi superfluids with different finite pairing interactions on both sides to extract the penetration depth, so there is no quantum critical point in real space like our setup. Moreover, having multiple superfluid phases in one setup may need one of them to be beyond the BCS regime and cause complications before the BCS behavior is thoroughly investigated.

\subsection{Spatial quench and spatial KZ mechanism}
On the other hand, to investigate the spatial KZ mechanism, we consider a more general type of quench of the pairing interaction. For a spatial quench,
\begin{equation}
    g'(x)= 
\begin{cases}
    c, &   0 \leq x/L < x_c/L-d, \\
    -\frac{c}{d}\frac{(x-x_c)}{L}, & (x_c/L-d)\leq x/L \leq x_c/L,\\
    0,              &  1>x/L>x_c/L.
\end{cases}
\end{equation}
Again, we typically set $x_c=L/2$, where the order parameter vanishes. Here $c$ and $d$ are dimensionless parameters. $-c/d$ is the slope of the linear ramp shown in Fig. \ref{fig:gff}.

In the spatial KZM, the freezing-out of the correlation length within the linear-ramp regime is the key to extract the scaling behavior of the correlation length. Explicitly, one considers a dimensionless parameter $\epsilon$ to identify the distance to the critical point, which occurs at $x_c$ separating the two phases, with the relation 
\begin{equation}
    \epsilon(x) = \alpha (x-x_c).
\end{equation}
We choose $x<x_c$ to represent the broken-symmetry (superfluid) phase and $x>x_c$ to represent the symmetric (normal-gas) phase.  
For a typical second-order phase transition in a uniform system, the correlation length diverges according to $\xi\sim \epsilon^{-\nu}$ near the critical point. For the spatial quench, the critical point is at $x_c$ in real space, so the local correlation length diverges as $\xi\approx (\alpha|x_c-x|)^{-\nu}$~\cite{Zurekptis}. Within a distance $|x_h-x_c|$ from $x_c$, the correlation length reaches the same order as the distance: $|x_h-x_c|\approx (\alpha|x_c-x_h|)^{-\nu}$. This sets a frozen correlation length of 
$\xi \sim \alpha^{-\frac{\nu}{1+\nu}}$.
Thus, the spatial KZM predicts that the penetration into the symmetric phase decays with a characteristic length $\xi$.

However, the zero-temperature BCS theory near $g=0$ does not feature a power-law divergence of $\xi$. 
The BCS coherence length is~\cite{Fetter_book,Leggett} 
\begin{equation}\label{eq:BCSxi}
\xi_{\Delta}=\frac{\hbar^2 k_f}{ m \Delta}. 
\end{equation}
The Fermi momentum is related to the local density  via $k_f=\pi \rho/2$ in 1D. 
In the weakly interacting limit, the gap function at zero temperature is given by \cite{Fetter_book,Pethick-BEC}
\begin{equation}\label{eq:BCS-delta}
    \Delta=\frac{8}{e^2}~E_f~ e^{-1/\mathcal{N}g},
\end{equation}
where $\mathcal{N}=\frac{m}{\pi\hbar^2k_f}$ is the density of states at the Fermi energy in 1D.
Therefore, the BCS coherence length in the weakly interacting limit ($g\rightarrow 0$) becomes
\begin{equation}\label{eq:BCSxi-theory}
    \xi_{\Delta}=\frac{e^2}{4k_f}e^{1/\mathcal{N}g}.
\end{equation}
We caution that the expression is non-analytic in $g$.
To study the spatial quench, we identify $\alpha=\frac{c}{dL}$, so
$    g'(x)=-\alpha(x-x_c)$
in the ramp-down region and remark that the sign convention does not affect the scaling analysis.
The frozen-out correlation $\xi_{fr}$ occurs when Eq.~\eqref{eq:BCSxi-theory} is met by  $g'=\alpha\xi_{fr}$, so
$\xi_{fr}\sim e^{k_f/\mathcal{N}\alpha\xi_{fr}E_f}$.
After simplifying the expression with dimensionless quantities, such as $\tilde{\xi}_{fr}=\xi_{fr}/L$, we obtain
\begin{equation}\label{eq:BCSKZM}
    f(\tilde\xi_{fr}) \equiv \frac{\tilde\xi_{fr}}{2\pi}\ln(\frac{4k_fL}{e^2}\tilde\xi_{fr})\sim \frac{1}{\alpha L}.
\end{equation}
Thus, the spatial KZM for Fermi superfluid at zero temperature has the above form due to the non-analytic behavior of the $T=0$ BCS theory.
To better contrast the mechanisms and features of the step-function and spatial quenches, we compare them with the corresponding continuous phase transition of a uniform system in Table~\ref{tab:Comparison}.

\begin{table}[t]
    \centering
    \begin{tabular}{|l|l|l|l|}
    \hline
         & Continuous & Step-function  & Spatial quench  \\
         & phase transition & quench & (spatial KZM) \\
         \hline
         Parameter & uniform & sudden drop & linear ramp \\
         \hline
         Structure & uniform & coexistence & coexistence \\
         \hline
         Transition & whole system & $x=x_c$ & $x=x_c$ \\
         \hline
         Penetration & N/A & Correlations & Correlations \\
         \hline
         Power law & $\xi\sim \epsilon^{-\nu}$ & N/A & $\xi\sim\alpha^{-\frac{\nu}{1+\nu}}$ \\
         \hline
         BCS ($T=0$) & $\xi\sim e^{1/\mathcal{N}g}$ & $\xi\sim \hbar v_{F0}/\Delta_0$ & $\xi \ln(\xi)\sim 1/\alpha$ \\
         \hline
    \end{tabular}
    \caption{Comparison of continuous phase transition in a homogeneous system, step-function quench, and spatial quench described by the spatial KZM. Here $x_c$ is the location where the parameter drops to zero, separating the symmetric and symmetry-broken phases in real space, $\xi$ is the correlation length, $\epsilon$ is the distance to the critical point, $\alpha$ is the slope of the parameter ramp, $\mathcal{N}$ is the density of state at the Fermi energy, $v_{F0}$ and $\Delta_0$ are the bulk Fermi velocity and gap on the superfluid side. We emphasize all three cases are in equilibrium.}
    \label{tab:Comparison}
\end{table}

\section{Results and discussions}\label{sec:result}

\subsection{Numerical calculations}
To solve the BdG equation, we begin with  chemical potential $\mu$ and an initial trial for $\Delta(x)$ and find the eigenvalues and eigenfunctions from the BdG equation.
We then assemble $\Delta(x)$ from the eigenfunctions using Eq.~\eqref{Eq:DeltaBdG}. 
The new gap function is used in the BdG equation to find the new eigenvalues and eigenfunctions. We continue the iteration until the consistency condition $\int |\Delta^{old}-\Delta^{new}|dx<10^{-5}$ is met.
We then adjust $\mu$ and repeat the above steps until we meet the condition $N=\int\rho(x) dx$ using Eq.~\eqref{Eq:rhoBdG}.
The number of grid points $n_x$ to discretize the real space imposes a momentum cutoff $k_{max}=\frac{\pi n_x}{2L}$. We choose $n_x$ large enough that the results are insensitive to further changes of $n_x$. Most of our calculations are for half filling with $n_x=N$. The results not far away from half filling are qualitatively the same.  However, physical quantities may have relatively large fluctuations far way from half filling due to the small ratio of $\Delta(x)/E_f^0$.
We have verified that for uniform BCS superfluid,  the BdG results from our calculations reproduce the known results in the literature~\cite{Fetter_book,Pethick-BEC}.

In both step-function and spatial quenches, $\Delta(x)$ drops to zero when $g(x)=0$ according to Eq.~\eqref{Eq:DeltaBdG}. However, the pair wavefunction $F(x)$ can penetrate into the normal region with $g(x)=0$. 
We will analyze the penetration in different settings and characterize the correlation length $\xi$.
The correlation function on the noninteracting side according to Eq. (\ref{eq:correlation}) can be evaluated by
\begin{equation}
        C(r)=\frac{1}{n_x-r'}\sum_{n+r'\leq n_x}F(x_n)F(x_{n+r'}),
    \end{equation}
    where $r=r' dx$, 
    $n={1,\cdots,n_x}$ and
    $r'={1,\cdots,n_x/2}$ are integers.

To extract the scaling behavior from the quench protocols, we fit $F(x)$ in the noninteracting region by the exponential form \eqref{eq:exp-decay} and the power-law form \eqref{eq:power-law}. 
As expected, the power-law fits $F(x)$ better in both step-function and spatial quenches. However, the exponential form may produce similar exponents even though the fitting does not faithfully go through the data. On the other hand, fitting the pair-pair correlation function $C(r)$ with a power-law similar to Eq.~\eqref{eq:power-law} results in significant deviations in both step-function and spatial quenches, but $C(r)$ can be fitted reasonably well with the exponential function $C_0 \exp(-r/\xi_C)$. 
We extract the correlation lengths from $F(x)$ and $C(r)$ and denote them by $\xi_F$ and $\xi_C$, respectively, and  introduce the dimensionless quantities $\tilde{\xi}_{C,F}=\xi_{C,F}/L$.

We also evaluated the BCS coherence length defined in Eq.~\eqref{eq:BCSxi} by using the bulk values on the superfluid side. In general, the evaluation of $\xi_{\Delta}$ becomes less reliable when the bulk $\Delta$ suffers strong fluctuations in the weakly interacting regime with $c<1$. On the other hand, there are also restrictions on the fitting of $F(x)$ and $C(r)$, as will be explained below. In our analysis, we stay within the reliable regimes for extracting the scaling behavior. 

\begin{figure}[t]
    \centering
    \includegraphics[width=0.9\columnwidth,keepaspectratio]{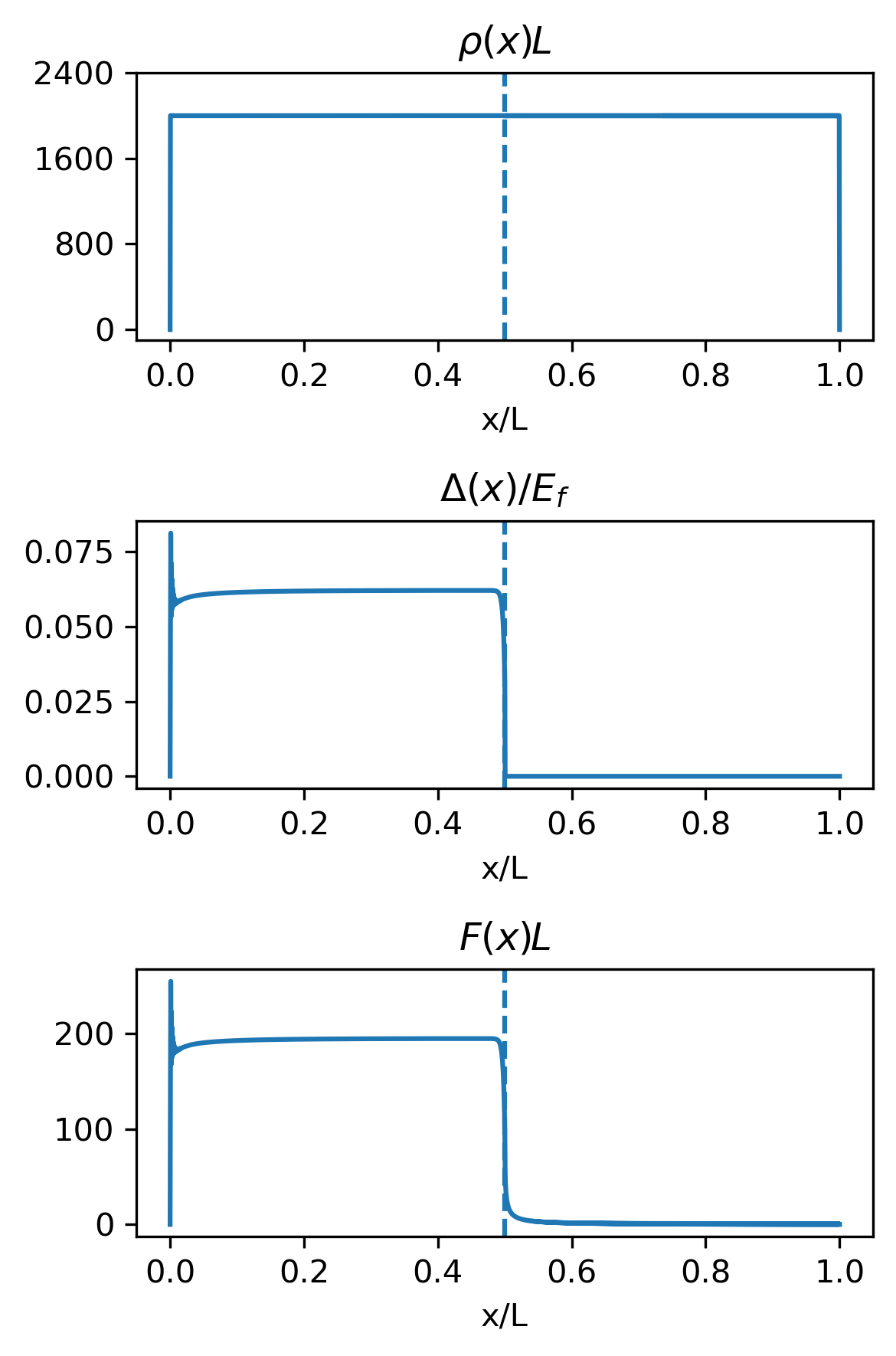}
    \caption{Profiles of the density (top), gap function (middle), and pair wavefunction (bottom) in a step-function quench. The vertical dashed lines indicate where the pairing interaction drops to zero. Here $n_x=2000$, $N=2000$, and $c=1$.
    }
    \label{fig:sfq-plots}
\end{figure}

\begin{figure}[t]
        \centering
        \includegraphics[width=0.95\columnwidth]{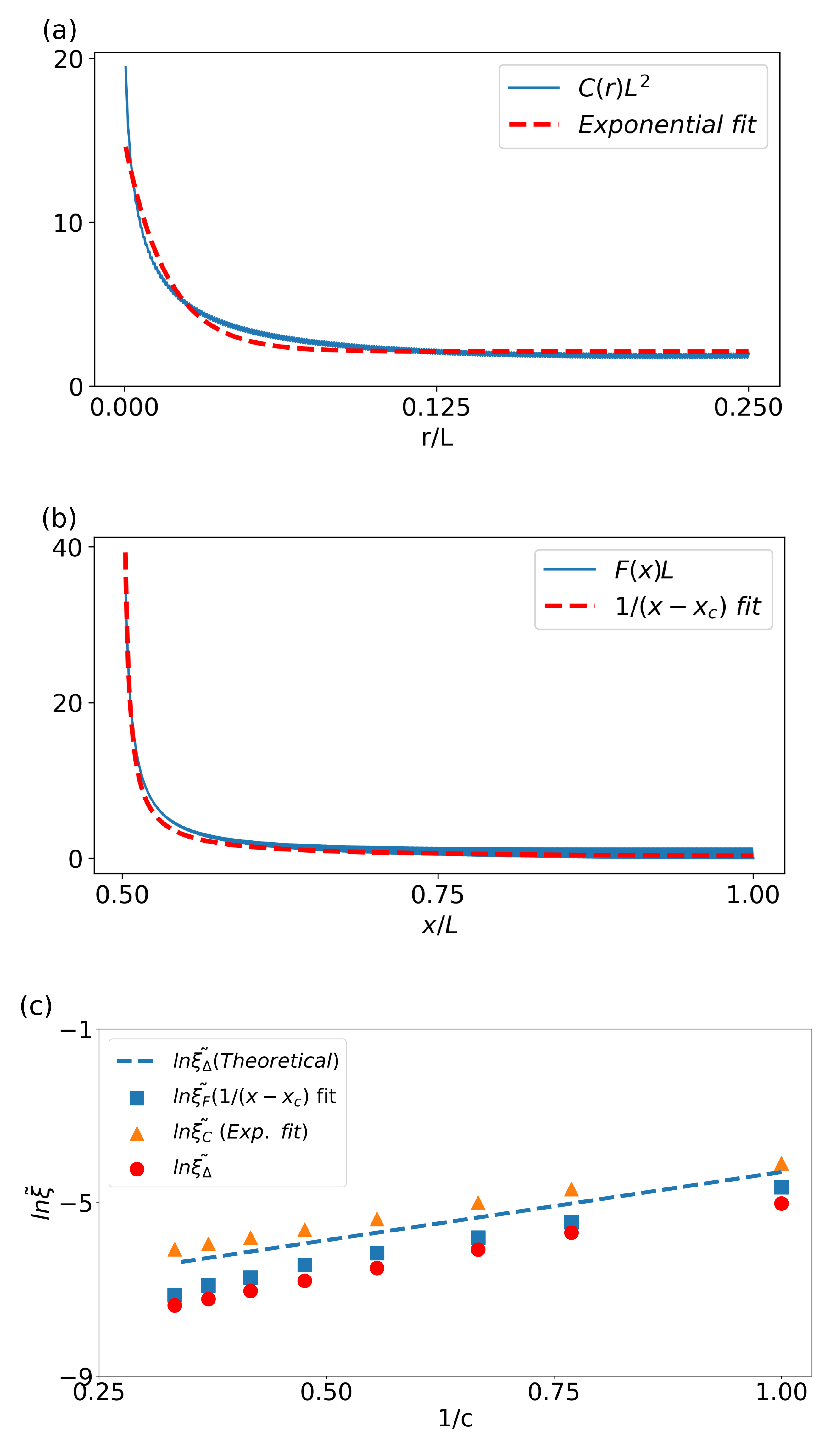}
        \caption{Correlation lengths in the step-function quench. (a) The pair correlation function $C(r)$ (solid line) and it exponential fit (dashed line). (b) The pair wavefunction $F(x)$ (solid line) and its power-law fit (dashed line). (c) Scaling behavior with respect to $1/c$ of $\tilde{\xi}_F=\xi_F/L$ from the power-law fit (squares),  $\tilde{\xi}_C=\xi_C/L$ from the exponential fit (triangles), and the BCS coherence length $\xi_\Delta=\tilde{\xi}_\Delta L$ (circles). The dashed  line represents the BCS approximation of the coherence length at zero temperature given by Eq. (\ref{eq:BCSxi-theory}). Here $n_x=2000$ and $N=2000$. In (a) and (b), $c=1$.
        \label{fig:curvefit}      
        }
    \end{figure}

\subsection{Step-function quench}
As shown in Fig.~\ref{fig:sfq-plots}, though the density profile is basically uniform inside the box in the presence of a step-function quench, the order parameter vanishes at the critical point in real space. 
For the step-function quench, the correlation lengths $\xi_F$ and $\xi_C$ along with their fitting curves and the BCS coherence length $\xi_{\Delta}$ are shown in Fig. \ref{fig:curvefit}. The scaling behavior allows us to extract their exponents. However, the range of $c$ is limited for $\xi_F$ and $\xi_C$ because if $c<1$, the gap function $\Delta(x)$ is small and suffers strong fluctuations in the superfluid region. If $c>3$, the correlation lengths $\xi_F$ and $\xi_C$ may go below the numerical resolution, and the fitting also shows observable deviations. 

As suggested in the studies of proximity effects in SC-NM junctions \cite{degennes-proximity,Falk-PE,PE-Silvert}, the dominant length scale in the penetration of Cooper pairs is the BCS coherence length $\xi_\Delta$.
Increasing the pairing interaction leads to stronger binding between the fermions, which results in a smaller BCS coherence length as the pairs are more tightly bound in real space. One can also see that increasing the pairing interaction increases the bulk $\Delta$ and decreases the BCS coherence length according to Eq.~\eqref{eq:BCSxi}.

Fig. ~\ref{fig:curvefit} (c) shows that the correlation lengths $\xi_F$ and $\xi_C$ and the BCS coherence length $\xi_{\Delta}$ from the step-function quench 
all exhibit the same scaling behavior of Eq.~\eqref{eq:BCSxi-theory} in . 
Hence, our results support the proposition that the correlation lengths $\xi_F$ and $\xi_C$ follow $\xi_\Delta$, so the correlation lengths decrease with the BCS coherence length as the pairing interaction increases. Our results also confirm that $\xi_\Delta$ from the superfluid region may be considered as the only relevant length scale besides the box size $L$ in a step-function quench. 
The chemical potential in the study of the step-function quench is about $\mu \sim 0.9 E_f^0$, indicating the system is still in the BCS regime. Moreover, the correlation and coherence length follow the BCS coherence length in the weakly interacting limit, as shown in Fig.~\ref{fig:curvefit}.  Hence, we have presented a fair comparison of the different coherence and correlation lengths in the step-function quench.

 


\begin{figure}[t]
    \centering
    \includegraphics[width=0.9\columnwidth,keepaspectratio]{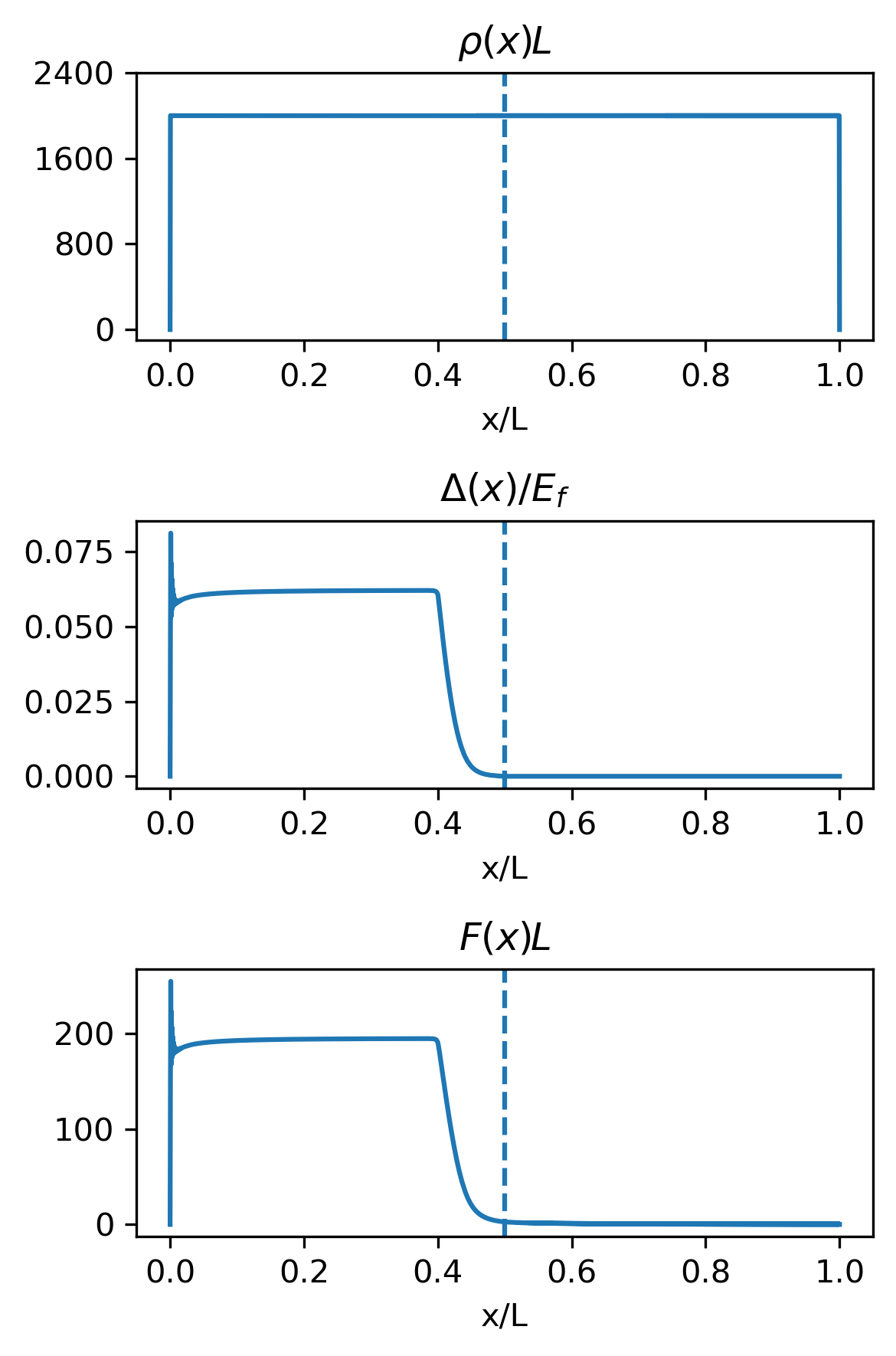}
    \caption{Profiles of the density (top), gap function (middle) and pair wavefunction (bottom) in a spatial quench. The vertical dashed lines indicate where the interaction drops to zero. Here $n_x=2000$, $N=2000$, $c=1$, and $d=0.1$.
    } 
    \label{fig:sq-plots}
\end{figure}

\subsection{Spatial quench}
\begin{figure}[t]
    \centering
    \includegraphics[width=0.9\columnwidth,keepaspectratio]{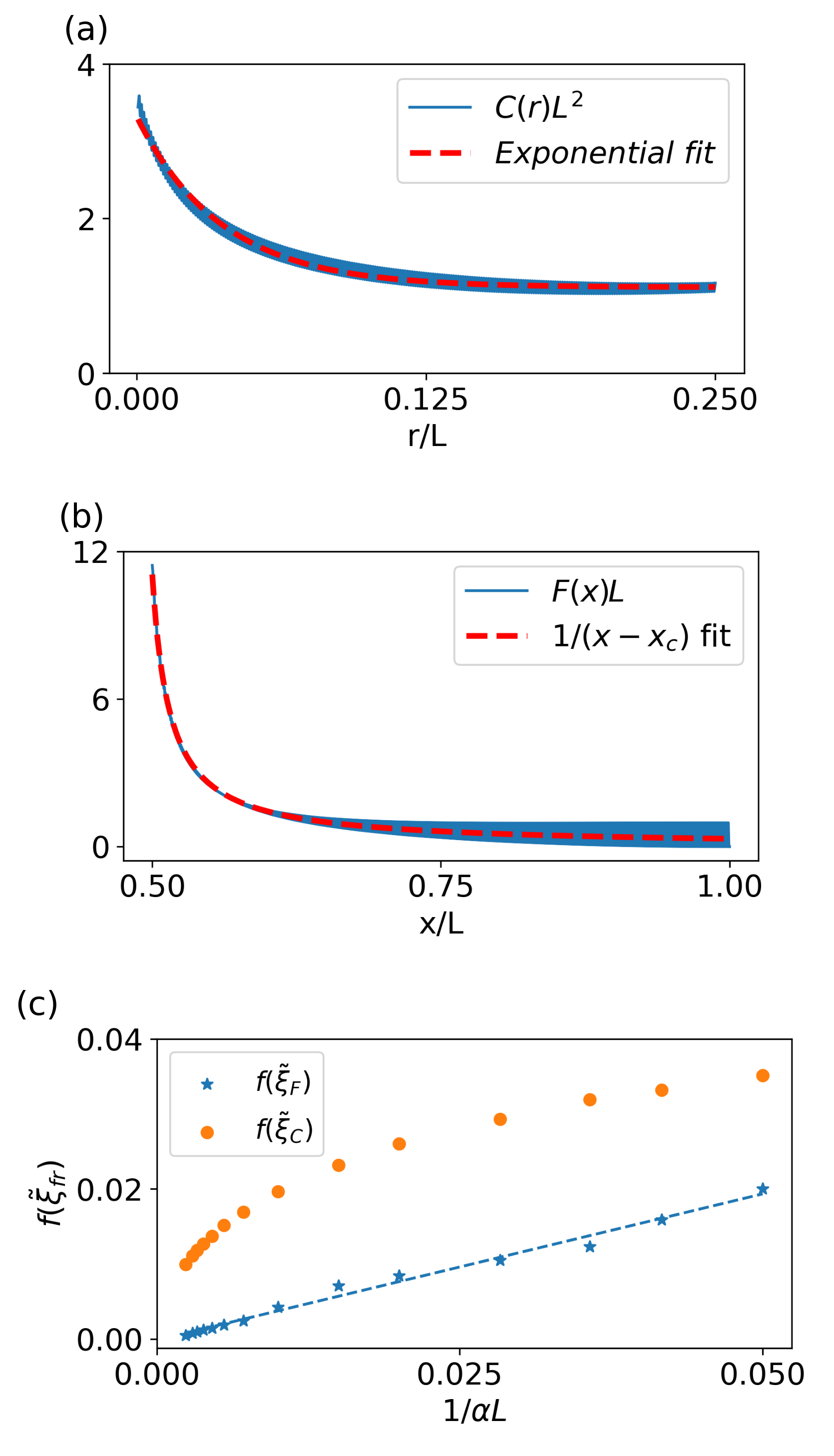}
    \caption{Correlation lengths  in the spatial quench. (a) $C(r)$ (solid line) and its exponential fit (dashed line). (b) $F(x)$ (solid line) and its power-law fit (dashed line). Here $n_x=2000$,  $N=2000$, $c=5$, and $d=0.1$. (c) Scaling behavior according to Eq.~\eqref{eq:BCSKZM} with respect to $1/(\alpha L)$ for  $\tilde{\xi}_F=\xi_F/L$ (starts) and $\tilde{\xi}_C=\xi_C/L$ (circles). The dashed line is a linear fit to $f(\tilde{\xi_F})$. Here $\alpha=c/(dL)$, $n_x=2000$, and $N=2000$.
    }
    \label{fig:sq-scaling}
\end{figure}

For the spatial quench, the pairing interaction ramps down linear from the superfluid region to zero in the normal-gas region within a distance $d$. Figure~\ref{fig:sq-plots} shows the profiles of density, order parameter $\Delta$, and pair wavefunction $F$ for a selective case of spatial quench. The linear-ramp region of the interaction leads to more complicated behavior between the bulks of the superfluid and normal gas.
For $F(x)$, the power-law form \eqref{eq:power-law} again fits the penetration better, but the exponential form \eqref{eq:exp-decay} gives close answers despite more significant deviations. In contrast, the power-law form cannot reasonably fit to $C(r)$ in the normal-gas regime while the exponential form $C_0 \exp(-r/\xi_C)$ fits reasonably well, as shown in Fig.~\ref{fig:sq-scaling} (a) and (b).

After extracting the correlation lengths $\xi_F$ and $\xi_C$ from the fitting, their scaling behavior according to Eq.~\eqref{eq:BCSKZM} are analyzed in Fig.~\ref{fig:sq-scaling} (c). Despite the non-analytic behavior of $f(\xi)$, the correlation length $\xi_F$ from the pair wavefunction follows the relation~\eqref{eq:BCSKZM}, as the linear fit on the plot suggests. In contrast, the correlation length $\xi_C$ from the correlation function exhibits observable deviations from the scaling behavior of Eq.~\eqref{eq:BCSKZM}, possibly due to higher-order correlations. Therefore, the spatial quench of Fermi superfluid differentiates the correlation lengths $\xi_F$ and $\xi_C$ from the BdG equation, and the correlation length $\xi_F$ follows the scaling behavior predicted by the spatial KZM according to the mean-field BCS theory.


Different from the step-function quench, here we have a larger window to check scaling of the correlation lengths with respect to the slope $\alpha$ for the spatial quench. Moreover, we have checked the scaling behavior of the correlation lengths independently for the parameters $c$ and $d$ and confirmed the consistency of the scaling with respect to $\alpha$. 
For the range of $\alpha$ tested in our study, the chemical potential is around $\mu\sim0.9 E_f^0$, again indicating the system is in the BCS regime with half filling. However, the chemical potential can change for lower filling as $\alpha$ changes.
The density change that affects $k_f$ is virtually non-observable as $d$ changes in our study.

\begin{figure}[t]
    \centering
    \includegraphics[width=0.7\columnwidth,keepaspectratio]{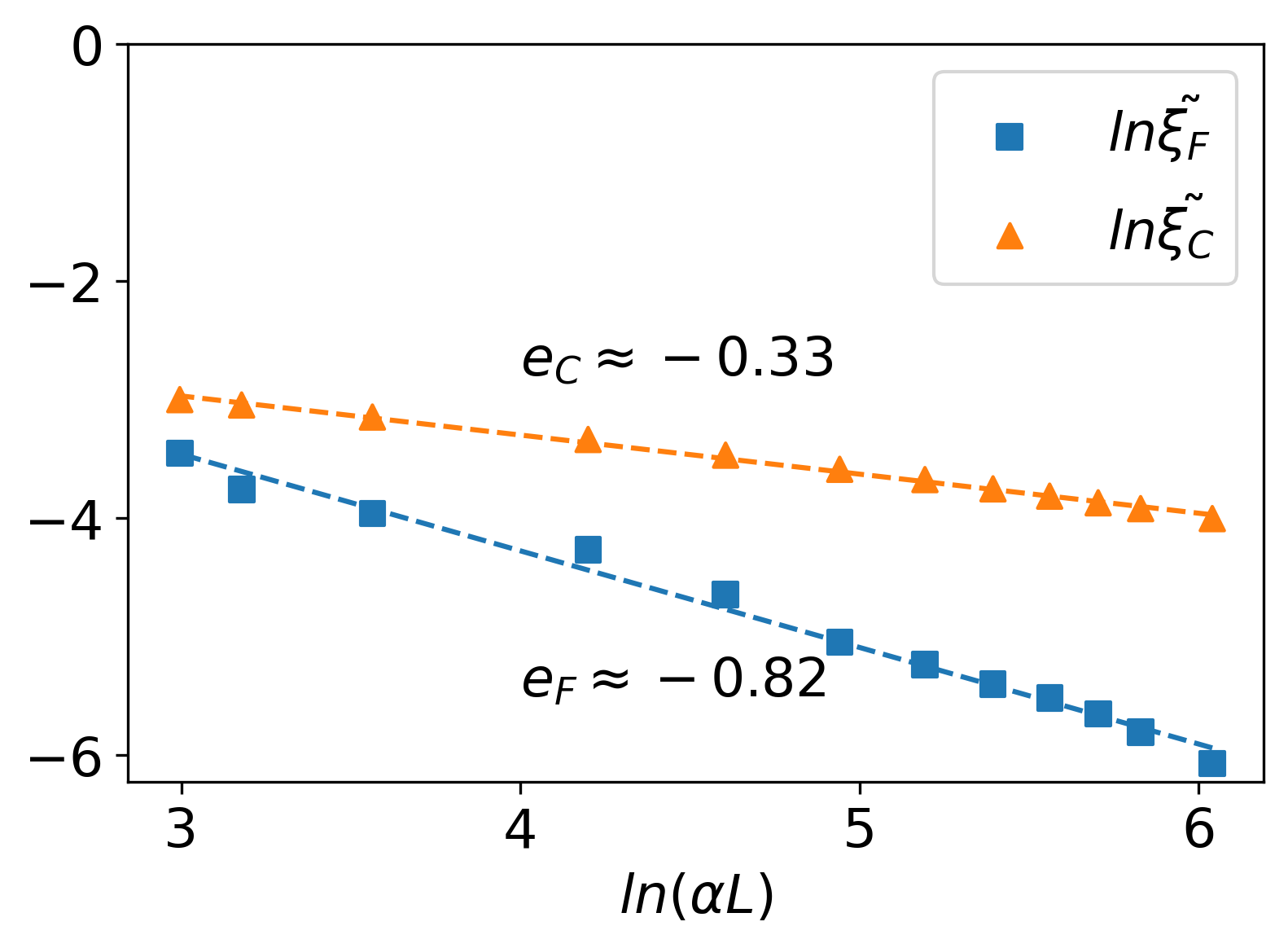}
    \caption{Local power-law scaling behavior with respect to $\alpha L$ for  $\tilde{\xi}_F=\xi_F/L$ (squares) and $\tilde{\xi}_C=\xi_C/L$ (triangles). Here $n_x=2000$ and  $N=2000$. The lines are power-law fits with the exponents labeled next to the data.}
    \label{fig:sq-xi-vs-alpha}
\end{figure}

We mention that for the range of $\alpha$ that we tested in spatial quench, the correlation lengths may mimic the power-law scaling with respect to $\alpha$. As shown in Fig.~\ref{fig:sq-xi-vs-alpha}, both $\xi_F$ and $\xi_C$ can be locally fitted by a power law and extract the corresponding exponent.
We found the exponent from $\xi_C$ is close to $-1/3$ but that from $\xi_F$ is more than twice larger. While this local analysis of power-law behavior again shows that the spatial KZM of Fermi superfluid in the BCS framework indeed differentiates the correlation lengths from the pair wavefunction and its correlation function, the non-analytic behavior of the BCS theory at $T=0$ leading to Eq.~\eqref{eq:BCSKZM} shows that Fig.~\ref{fig:sq-scaling} (c) captures the full scaling of the correlation lengths while Fig.~\ref{fig:sq-xi-vs-alpha} only shows how the non-analytic behavior may disguise itself as power-law behavior in a local analysis. We also remark that the Ginzburg-Landau theory of Fermi superfluid~\cite{Fetter_book} only works near the transition temperature, which may not apply to our analysis of the $T=0$ results.


\subsection{Bosonic background}
After discussing the step-function and spatial quenches of Fermi gases, we consider the quenches in the presence of a uniform bosonic background, which may come from sympathetic cooling ~\cite{Onofrio_2016} or boson-fermion superfluid mixtures~\cite{doi:10.1126/science.1255380}. In a simple setting, we consider fermions with two components and bosons in the same quasi-1D box of length $L$. There is attraction between fermions with opposite spins but repulsion between bosons and between fermions and bosons. As a first attempt to address the mixture, we only consider the inhomogeneous pairing interaction $g(x)$ between the fermions while keeping the other parameters uniform. By using the fermionic parameters as units, the boson-boson and boson-fermion coupling constants can be written in terms of dimensionless quantities as $g_{bb}=g'_{bb}E_f^0/k_f^0$ and $g_{bf}=g'_{bf}E_f^0/ k_f^0$, respectively.




Previous studies \cite{Tom-18,BFpaper} have shown that bosons and fermions in a binary mixture can form miscible mixtures when the inter-species interaction is relatively weak or the densities are low. However, phase-separation structures with inhomogeneous densities start to emerge as the inter-species interaction and densities increase. Moreover, the pressure of bosons is mainly from the boson-boson interactions, which competes with the Fermi pressure of the fermions.
Since we focus on the impact of the bosonic background on the quenches of fermions, we concentrate on the regime when the mixture is in the miscible phase. 
Instead of a full analysis of various atomic boson-fermion mixtures, we
check a specific case of $^7$Li - $^6$Li mixtures with equal population of all species. The conditions $g'_{bf}<<g'_{bb}$ and half-filling are sufficient to maintain a miscible phase for the selected case. However, the formalism presented here is generic and can be applied to atomic boson-fermion mixtures in general.

The total ground-state energy functional of a mixture of bosons and fermions in a quasi-1D box of length $L$, assuming the fermions form a BCS superfluid, is given by
\begin{equation}\label{eq;Emix}   E_{mix}=E_g+E_b+g_{bf}\int_0^L dx \rho_b(x)\rho(x).
\end{equation}
Here, $E_g$ is the BCS ground-state energy shown in Eq.~\eqref{Eq:HBdG},
and the energy of the bosons is
\begin{equation}
   E_b=\int_0^L dx [\frac{\hbar^2}{2m_b}|\partial_x\psi_b|^2
+\frac{1}{2}g_{bb}|\psi_b|^4].
\end{equation}
In the mean-field description of the ground state, the condensate wavefunction of the bosons is governed by the Gross-Pitaevskii (GP) equation \cite{Pethick-BEC,pitaevskii2003bose}. To find the minimal-energy configuration, we implement the imaginary-time formalism~\cite{Fetter_book,Pethick-BEC} by searching for the stable solution to the imaginary-time evolution equation $-\partial \psi_b/\partial \tau=\delta E_{mix}/\delta \psi_b^*$ in the $\tau\rightarrow \infty$ limit, starting from a trial initial configuration.  The normalization $\int|\psi_b|^2 dx =N_b$ is imposed at each imaginary-time increment to project out higher-energy states. Here $\tau=it$ is the imaginary time. Explicitly,
\begin{equation}\label{gpeqn1}
       -\hbar\frac{\partial \psi_b}{\partial \tau} = -\frac{\hbar^2}{2m_b}\partial_x^2\psi_b +   g_{bb}\rho_b\psi_b
      + g_{bf}\rho\psi_b,
\end{equation}
The fermions are described by the BdG equation~\eqref{discrete-BdG} with the replacement of the discretization of $h(x)=-\frac{\hbar^2}{2m}\frac{\partial^2}{\partial x^2}-\mu+g_{bf}\rho_b$. The bosonic density is $\rho_b(x)=|\psi_b(x)|^2$ while $\rho(x)=2\sum_{\tilde{n}}'|v_{\tilde{n}}(x)|^2$ for the fermions as before.

For a uniform and miscible mixture of bosons and fermions, the mean-field treatment shifts the chemical potential of the fermions by $g_{bf}\rho_b$, which only shows up in the diagonal of the BdG equation. Therefore, the gap function is not affected directly by the bosons. This implies that the scaling of the fermionic correlation functions are insensitive to the bosonic background as long as the mixture remains uniform and miscible. However, the presence of step-function or spatial quench of boson-fermion mixtures in a quasi-1D box may introduce complications due to the inhomogeneous pairing interaction and confining potential. We numerically solve the coupled BdG and GP equations for a miscible boson-fermion mixtures in a quasi-1D box to verify if the exponents of the fermionic correlation lengths $\xi_F$ and $\xi_C$ are affected by the bosonic background.

To solve the coupled BdG and GP equations by self-consistent iteration with given numbers of the bosons $N_b$ and fermions $N$, we begin with trial chemical potential $\mu$, boson wavefunction  $\psi_b$, and gap function $\Delta(x)$ and first solve the BdG equation following the procedure implemented in the previous sections. The gap function $\Delta(x)$ and fermionic density $\rho(x)$ are then obtained from the eigenfunctions $u_{\tilde{n}}(x)$ and $v_{\tilde{n}}(x)$.
Next, we evolve the imaginary-time evolution equation (\ref{gpeqn1}) to get the ground-state bosonic density $\rho_b(x)$. We continue the iterations between the BdG and GP equations until the final convergence of the gap function $\Delta(x)$ and the bosonic density $\int_0^L|\rho_b^{old}(x)-\rho_b^{new}(x)|dx<10^{-5}$ is reached.

\begin{figure}[t]
\centering
\includegraphics[width=0.99\columnwidth]{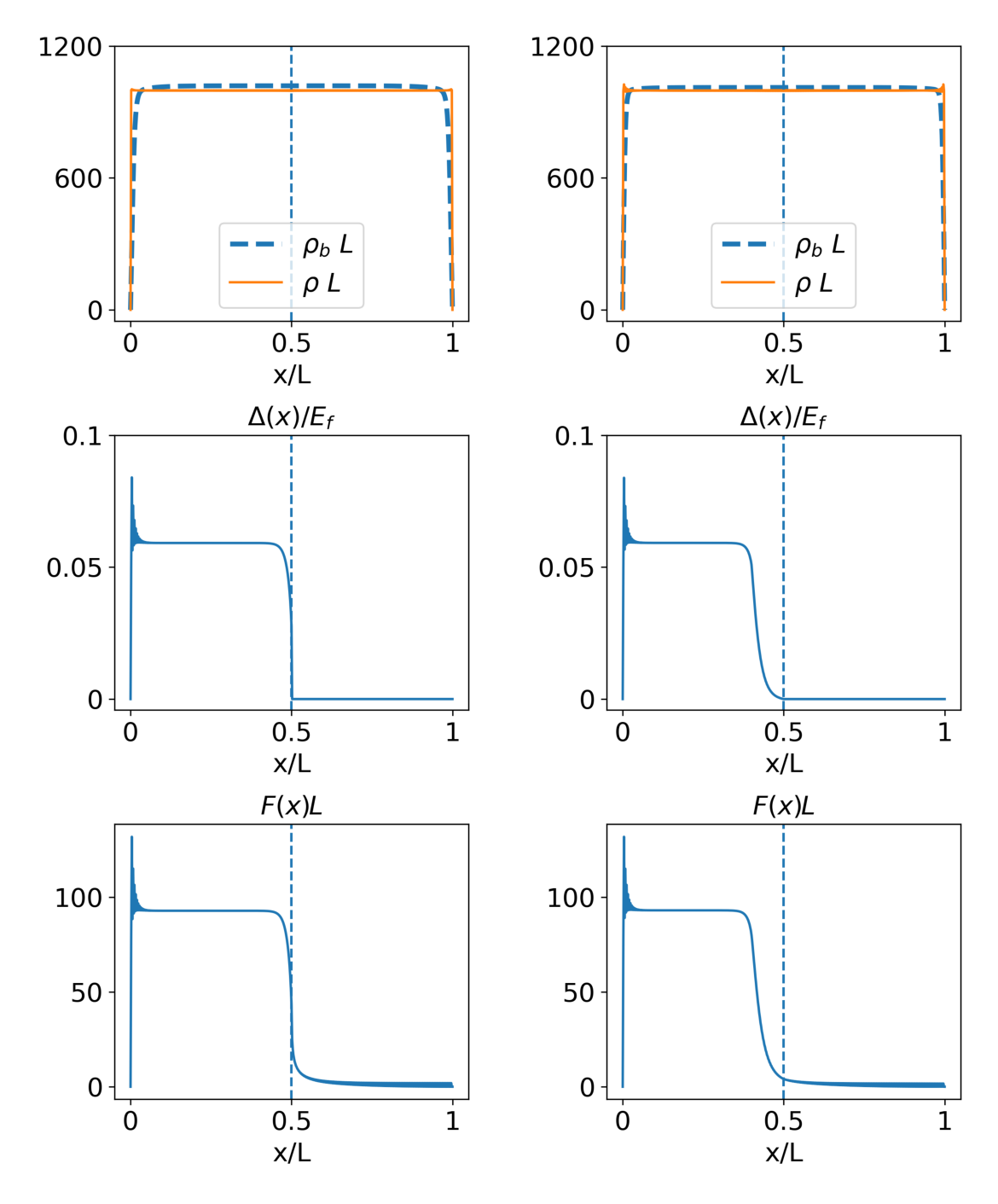}
    \caption{Profiles of the density (top row), gap function (middle row), and pair wavefunction (bottom row) of the step-function quench (left column) and spatial quench (right column) of $^6$Li in a $^6$Li-$^7$Li mixture. The inhomogeneous interaction only applies to the pairing interaction between the fermions, and other interactions are uniform. The vertical dashed lines indicate where the interaction drops to zero. For the step-function quench, $n_x=1000$, $N=N_b=1000$, and $c=1$. For the spatial quench, $n_x=1000$, $N=N_b=1000$, $c=1$, and $d=0.1$. For both cases, $g'_{bb}=0.1$ and $g'_{bf}=0.05$.
    }
    \label{fig:my_label}
\end{figure}

During the iterations, we also adjust the chemical potential $\mu$  for the BdG equation to meet the fixed number of total fermions.
Similar to the case with only fermions, different initial states have been used to confirm the ground state for both species by checking the ground-state energy using Eq.~\eqref{eq;Emix}. 
Since we focus on the case with a uniform bosonic background, we confine our parameters to $g'_{bf}<<g'_{bb}$, where the convergence to the miscible phase is found in all our trials of the initial states.
Similar to the procedures of step-function and spatial quenches discussed above, we calculated the pair wavefunction $F(x)$ and correlation function $C(r)$ to extract the corresponding correlation lengths $\xi_F$ and $\xi_C$, respectively.

Samples of the profiles of the density, gap function, and pair wavefunction of the step-function and spatial quenches of fermions in a boson-fermion mixture are shown in Fig.~\ref{fig:my_label}.
For the step-function quench, we also evaluate the BCS coherence length $\xi_\Delta$ from the bulk values on the superfluid side. From our numerical results, we found that the inclusion of a bosonic background with uniform parameters does not alter the scaling behavior of $\xi_F$ and $\xi_C$ of the fermions. All the scaling from the step-function and the exponents from the spatial quenches of boson-fermion mixtures are within numerical accuracy the same as those without the bosons, which have been shown in Figs.~\ref{fig:curvefit} (c) and ~\ref{fig:sq-scaling}. As shown in Fig.~\ref{fig:my_label}, this is mainly because the density profile of bosons becomes quite flat already at relatively small $g'_{bb}$, making the bosonic background basically uniform and does not further complicate the behavior of the fermions.

Nevertheless, the phase-separation structures of boson-fermion mixtures can exhibit various inhomogeneous profiles already for binary mixtures in the presence of uniform interactions~\cite{Tom-18,BFpaper}. Adding inhomogeneous interactions to the fermions, such as the step-function or spatial quench of the pairing interaction studied here, is expected to lead to richer structures. Extracting the correlation lengths in such highly inhomogeneous setups will be a challenge and await future research.

\section{Discussion and Implication}\label{sec:implication}
We elaborate on some subtle differences between the spatial KZM in the transverse field Ising model studied in Refs. \cite{Dziarmaga-doiqpt,Zurekptis} and the quasi-1D Fermi gases studied here.
The absence of interaction in the normal-gas region of the two-component Fermi gas resembles the spatial quench of the magnetic field in the quantum Ising model~\cite{Zurekptis}, where the field is absent in the ferromagnetic phase. However, the broken-symmetry phase of the transverse-field Ising model is in the region without the magnetic field while the broken-symmetry phase of the Fermi gas is in the region with finite pairing interactions.

In the study of the spatial KZM of the quantum transverse field Ising model \cite{Dziarmaga-doiqpt,Zurekptis}, it was shown that both the magnetization, which is the expectation of the local spin and corresponds to the order parameter, and the spin-spin correlation function exhibit the same scaling behavior in the symmetric phase. The exponents extracted from both quantities agree with the spatial KZM prediction.
In contrast, we have shown that for the spatial KZM of Fermi gases with spatially varying pairing interactions, the scaling behavior of the correlation length from the the pair wavefunction $F(x)$ differs from that from the pair correlation function $C(r)$ because of the non-analytic behavior of the $T=0$ BCS theory and possible higher-order correlations.   
Therefore, the scaling behavior of the Fermi superfluid with spatial quench of the pairing interaction exhibits rich contents and extends the scope of the spatial KZM.

Experimentally, ultracold atoms have been usually subject to uniform interactions due to the small cloud size compared to the magnetic field for tuning  Feshbach resonance \cite{RevModPhys.82.1225,Pethick-BEC}. 
There have been several ways for inducing inhomogeneous interactions in cold atoms. One approach is to use optical techniques to control the interactions between atoms. Examples include optical Feshbach resonance~ \cite{theory-optical-feshbach1,exp-optical-feshbach1,exp-optical-feshbach2} and optically controlled magnetic Feshbach resonance \cite{exp-opt-mag-fesbach,PhysRevLett.115.155301}.
Refs.~\cite{PhysRevLett.115.155301,PhysRevLett.122.040405} demonstrate spatial modulation of the interaction in BEC~\cite{PhysRevLett.115.155301} and $^6$Li fermions~\cite{PhysRevLett.122.040405} by optical controls with high speed and precision. Optical techniques may suffer atom loss and heating, so they are more suitable for changing the interaction with short length or time scale. Another approach is based on magnetic Feshbach resonance and magnetic field gradient~\cite{PhysRevLett.125.183602}, which allows for longer observation time. Thus, the inhomogeneous interactions for realizing the step-function and spatial quenches may become feasible with the rapid developments in manipulating ultracold atoms. We also mention that two-component atomic Fermi gases in 3D \cite{PhysRevLett.118.123401} and 2D \cite{PhysRevLett.120.060402} box potentials have been realized, and similar techniques may realize atomic Fermi gases in quasi-1D box potentials in the future.

Recent progress in quantum gas microscopy allows mapping of site-resolved density- or spin- correlations of the Fermi Hubbard model \cite{MitraDebayan2018Qgmo,PhysRevLett.125.010403,PhysRevLett.125.113601}.
Ref. \cite{hartke2022direct} demonstrates site-resolved location and spin of each fermion in the attractive Fermi Hubbard system using a bilayer quantum-gas microscope and reveals the formation and spatial ordering of fermion pairs. In addition, radio-frequency (rf) spectroscopy has been used to measure the excitation energy that reveals the pairing gap in atomic Fermi gases \cite{zweirlein-2003,zweirlein-2007,zweirlein-2019-1,zweirlein-2019-2}. Future developments may allow spatial resolution of the rf spectroscopy for cold atoms.
Those spatially resolved measurements of the pairing correlation of atomic Fermi gases are promising for observing the scaling behavior of the step-function and spatial quenches analyzed here.

\section{Conclusion}\label{sec:conclusion}
We have shown that atomic Fermi gases with tunable interactions in real space provide a powerful simulator for studying the analogues of the proximity effect and spatial KZM.
Through numerical calculations with a step-function or spatial quench of the pairing interaction, we characterize the penetration of the pair wavefunction and pair correlation into the noninteracting region. The scaling analyses of the correlation lengths from the step-function and spatial quenches lead to the exponents of the corresponding quantities. For the step-function quench, the correlation lengths follow the BCS coherence length due to the lack of additional length scale in the system. In contrast, the correlation lengths of the pair wavefunction and pair correlation function exhibit different scaling behavior in the spatial quench.  
The rapid development in manipulating and measuring inhomogeneous structures of cold-atoms will allow us to explore more interesting phenomena in such a unified platform.

\begin{acknowledgments}
We thank Wojciech Zurek, Yan He, Chien-Te Wu, Shizeng Lin, and Chen-Lung Hung for stimulating discussions. This work was supported by the National Science Foundation under Grant No. PHY-2011360.
\end{acknowledgments}

\bibliographystyle{apsrev}

\begin{thebibliography}{99}
	\expandafter\ifx\csname natexlab\endcsname\relax\def\natexlab#1{#1}\fi
	\expandafter\ifx\csname bibnamefont\endcsname\relax
	\def\bibnamefont#1{#1}\fi
	\expandafter\ifx\csname bibfnamefont\endcsname\relax
	\def\bibfnamefont#1{#1}\fi
	\expandafter\ifx\csname citenamefont\endcsname\relax
	\def\citenamefont#1{#1}\fi
	\expandafter\ifx\csname url\endcsname\relax
	\def\url#1{\texttt{#1}}\fi
	\expandafter\ifx\csname urlprefix\endcsname\relax\def\urlprefix{URL }\fi
	\providecommand{\bibinfo}[2]{#2}
	\providecommand{\eprint}[2][]{\url{#2}}
	
	\bibitem[{\citenamefont{Pitaevskii and Stringari}(2003)}]{pitaevskii2003bose}
	\bibinfo{author}{\bibfnamefont{L.~P.} \bibnamefont{Pitaevskii}}
	\bibnamefont{and}
	\bibinfo{author}{\bibfnamefont{S.}~\bibnamefont{Stringari}},
	\emph{\bibinfo{title}{Bose-Einstein condensation}}, International series of
	monographs on physics ; 116 (\bibinfo{publisher}{Clarendon Press},
	\bibinfo{address}{Oxford, UK}, \bibinfo{year}{2003}), ISBN
	\bibinfo{isbn}{0198507194}.
	
	\bibitem[{\citenamefont{Pethick and Smith}(2008)}]{Pethick-BEC}
	\bibinfo{author}{\bibfnamefont{C.~J.} \bibnamefont{Pethick}} \bibnamefont{and}
	\bibinfo{author}{\bibfnamefont{H.}~\bibnamefont{Smith}},
	\emph{\bibinfo{title}{Bose–Einstein Condensation in Dilute Gases}}
	(\bibinfo{publisher}{Cambridge University Press}, \bibinfo{year}{2008}),
	\bibinfo{edition}{2nd} ed.
	
	\bibitem[{\citenamefont{Ueda}(2010)}]{Ueda-book}
	\bibinfo{author}{\bibfnamefont{M.}~\bibnamefont{Ueda}},
	\emph{\bibinfo{title}{Fundamentals and New Frontiers of Bose-Einstein
			Condensation}} (\bibinfo{publisher}{World Scientific},
	\bibinfo{address}{Singapore}, \bibinfo{year}{2010}),
	\eprint{https://www.worldscientific.com/doi/pdf/10.1142/7216},
	\urlprefix\url{https://www.worldscientific.com/doi/abs/10.1142/7216}.
	
	\bibitem[{\citenamefont{Fatemi et~al.}(2000)\citenamefont{Fatemi, Jones, and
			Lett}}]{exp-optical-feshbach1}
	\bibinfo{author}{\bibfnamefont{F.~K.} \bibnamefont{Fatemi}},
	\bibinfo{author}{\bibfnamefont{K.~M.} \bibnamefont{Jones}}, \bibnamefont{and}
	\bibinfo{author}{\bibfnamefont{P.~D.} \bibnamefont{Lett}},
	\bibinfo{journal}{Phys. Rev. Lett.} \textbf{\bibinfo{volume}{85}},
	\bibinfo{pages}{4462} (\bibinfo{year}{2000}),
	\urlprefix\url{https://link.aps.org/doi/10.1103/PhysRevLett.85.4462}.
	
	\bibitem[{\citenamefont{Theis et~al.}(2004)\citenamefont{Theis, Thalhammer,
			Winkler, Hellwig, Ruff, Grimm, and Denschlag}}]{exp-optical-feshbach2}
	\bibinfo{author}{\bibfnamefont{M.}~\bibnamefont{Theis}},
	\bibinfo{author}{\bibfnamefont{G.}~\bibnamefont{Thalhammer}},
	\bibinfo{author}{\bibfnamefont{K.}~\bibnamefont{Winkler}},
	\bibinfo{author}{\bibfnamefont{M.}~\bibnamefont{Hellwig}},
	\bibinfo{author}{\bibfnamefont{G.}~\bibnamefont{Ruff}},
	\bibinfo{author}{\bibfnamefont{R.}~\bibnamefont{Grimm}}, \bibnamefont{and}
	\bibinfo{author}{\bibfnamefont{J.~H.} \bibnamefont{Denschlag}},
	\bibinfo{journal}{Phys. Rev. Lett.} \textbf{\bibinfo{volume}{93}},
	\bibinfo{pages}{123001} (\bibinfo{year}{2004}),
	\urlprefix\url{https://link.aps.org/doi/10.1103/PhysRevLett.93.123001}.
	
	\bibitem[{\citenamefont{Theocharis et~al.}(2005)\citenamefont{Theocharis,
			Schmelcher, Kevrekidis, and Frantzeskakis}}]{inhomo-condensate1}
	\bibinfo{author}{\bibfnamefont{G.}~\bibnamefont{Theocharis}},
	\bibinfo{author}{\bibfnamefont{P.}~\bibnamefont{Schmelcher}},
	\bibinfo{author}{\bibfnamefont{P.~G.} \bibnamefont{Kevrekidis}},
	\bibnamefont{and} \bibinfo{author}{\bibfnamefont{D.~J.}
		\bibnamefont{Frantzeskakis}}, \bibinfo{journal}{Phys. Rev. A}
	\textbf{\bibinfo{volume}{72}}, \bibinfo{pages}{033614}
	(\bibinfo{year}{2005}),
	\urlprefix\url{https://link.aps.org/doi/10.1103/PhysRevA.72.033614}.
	
	\bibitem[{\citenamefont{Theocharis et~al.}(2006)\citenamefont{Theocharis,
			Schmelcher, Kevrekidis, and Frantzeskakis}}]{inhomo-condensate2}
	\bibinfo{author}{\bibfnamefont{G.}~\bibnamefont{Theocharis}},
	\bibinfo{author}{\bibfnamefont{P.}~\bibnamefont{Schmelcher}},
	\bibinfo{author}{\bibfnamefont{P.~G.} \bibnamefont{Kevrekidis}},
	\bibnamefont{and} \bibinfo{author}{\bibfnamefont{D.~J.}
		\bibnamefont{Frantzeskakis}}, \bibinfo{journal}{Phys. Rev. A}
	\textbf{\bibinfo{volume}{74}}, \bibinfo{pages}{053614}
	(\bibinfo{year}{2006}),
	\urlprefix\url{https://link.aps.org/doi/10.1103/PhysRevA.74.053614}.
	
	\bibitem[{\citenamefont{Niarchou et~al.}(2007)\citenamefont{Niarchou,
			Theocharis, Kevrekidis, Schmelcher, and Frantzeskakis}}]{inhomo-condensate3}
	\bibinfo{author}{\bibfnamefont{P.}~\bibnamefont{Niarchou}},
	\bibinfo{author}{\bibfnamefont{G.}~\bibnamefont{Theocharis}},
	\bibinfo{author}{\bibfnamefont{P.~G.} \bibnamefont{Kevrekidis}},
	\bibinfo{author}{\bibfnamefont{P.}~\bibnamefont{Schmelcher}},
	\bibnamefont{and} \bibinfo{author}{\bibfnamefont{D.~J.}
		\bibnamefont{Frantzeskakis}}, \bibinfo{journal}{Phys. Rev. A}
	\textbf{\bibinfo{volume}{76}}, \bibinfo{pages}{023615}
	(\bibinfo{year}{2007}),
	\urlprefix\url{https://link.aps.org/doi/10.1103/PhysRevA.76.023615}.
	
	\bibitem[{\citenamefont{Rodrigues et~al.}(2008)\citenamefont{Rodrigues,
			Kevrekidis, Porter, Frantzeskakis, Schmelcher, and
			Bishop}}]{inhomo-condensate4}
	\bibinfo{author}{\bibfnamefont{A.~S.} \bibnamefont{Rodrigues}},
	\bibinfo{author}{\bibfnamefont{P.~G.} \bibnamefont{Kevrekidis}},
	\bibinfo{author}{\bibfnamefont{M.~A.} \bibnamefont{Porter}},
	\bibinfo{author}{\bibfnamefont{D.~J.} \bibnamefont{Frantzeskakis}},
	\bibinfo{author}{\bibfnamefont{P.}~\bibnamefont{Schmelcher}},
	\bibnamefont{and} \bibinfo{author}{\bibfnamefont{A.~R.}
		\bibnamefont{Bishop}}, \bibinfo{journal}{Phys. Rev. A}
	\textbf{\bibinfo{volume}{78}}, \bibinfo{pages}{013611}
	(\bibinfo{year}{2008}),
	\urlprefix\url{https://link.aps.org/doi/10.1103/PhysRevA.78.013611}.
	
	\bibitem[{\citenamefont{Clark et~al.}(2015)\citenamefont{Clark, Ha, Xu, and
			Chin}}]{PhysRevLett.115.155301}
	\bibinfo{author}{\bibfnamefont{L.~W.} \bibnamefont{Clark}},
	\bibinfo{author}{\bibfnamefont{L.-C.} \bibnamefont{Ha}},
	\bibinfo{author}{\bibfnamefont{C.-Y.} \bibnamefont{Xu}}, \bibnamefont{and}
	\bibinfo{author}{\bibfnamefont{C.}~\bibnamefont{Chin}},
	\bibinfo{journal}{Phys. Rev. Lett.} \textbf{\bibinfo{volume}{115}},
	\bibinfo{pages}{155301} (\bibinfo{year}{2015}),
	\urlprefix\url{https://link.aps.org/doi/10.1103/PhysRevLett.115.155301}.
	
	\bibitem[{\citenamefont{Di~Carli et~al.}(2020)\citenamefont{Di~Carli,
			Henderson, Flannigan, Colquhoun, Mitchell, Oppo, Daley, Kuhr, and
			Haller}}]{PhysRevLett.125.183602}
	\bibinfo{author}{\bibfnamefont{A.}~\bibnamefont{Di~Carli}},
	\bibinfo{author}{\bibfnamefont{G.}~\bibnamefont{Henderson}},
	\bibinfo{author}{\bibfnamefont{S.}~\bibnamefont{Flannigan}},
	\bibinfo{author}{\bibfnamefont{C.~D.} \bibnamefont{Colquhoun}},
	\bibinfo{author}{\bibfnamefont{M.}~\bibnamefont{Mitchell}},
	\bibinfo{author}{\bibfnamefont{G.-L.} \bibnamefont{Oppo}},
	\bibinfo{author}{\bibfnamefont{A.~J.} \bibnamefont{Daley}},
	\bibinfo{author}{\bibfnamefont{S.}~\bibnamefont{Kuhr}}, \bibnamefont{and}
	\bibinfo{author}{\bibfnamefont{E.}~\bibnamefont{Haller}},
	\bibinfo{journal}{Phys. Rev. Lett.} \textbf{\bibinfo{volume}{125}},
	\bibinfo{pages}{183602} (\bibinfo{year}{2020}),
	\urlprefix\url{https://link.aps.org/doi/10.1103/PhysRevLett.125.183602}.
	
	\bibitem[{\citenamefont{Deutscher and de~Gennes}(1969)}]{degennes-proximity}
	\bibinfo{author}{\bibfnamefont{G.}~\bibnamefont{Deutscher}} \bibnamefont{and}
	\bibinfo{author}{\bibfnamefont{P.}~\bibnamefont{de~Gennes}},
	\bibinfo{journal}{pp 1005-34 of Superconductivity. Vols. 1 and 2. Parks, R.
		D. (ed.). New York, Marcel Dekker, Inc., 1969}  (\bibinfo{year}{1969}).
	
	\bibitem[{\citenamefont{Dziarmaga}(2010)}]{Dziarmaga-rev}
	\bibinfo{author}{\bibfnamefont{J.}~\bibnamefont{Dziarmaga}},
	\bibinfo{journal}{Adv. Phys.} \textbf{\bibinfo{volume}{59}},
	\bibinfo{pages}{1063} (\bibinfo{year}{2010}), ISSN \bibinfo{issn}{0001-8732}.
	
	\bibitem[{\citenamefont{Sachdev}(2011)}]{sachdev_2011}
	\bibinfo{author}{\bibfnamefont{S.}~\bibnamefont{Sachdev}},
	\emph{\bibinfo{title}{Quantum Phase Transitions}}
	(\bibinfo{publisher}{Cambridge University Press},
	\bibinfo{address}{Cambridge, UK}, \bibinfo{year}{2011}),
	\bibinfo{edition}{2nd} ed.
	
	\bibitem[{\citenamefont{Falk}(1963)}]{Falk-PE}
	\bibinfo{author}{\bibfnamefont{D.~S.} \bibnamefont{Falk}},
	\bibinfo{journal}{Phys. Rev.} \textbf{\bibinfo{volume}{132}},
	\bibinfo{pages}{1576} (\bibinfo{year}{1963}),
	\urlprefix\url{https://link.aps.org/doi/10.1103/PhysRev.132.1576}.
	
	\bibitem[{\citenamefont{Clark}(1968)}]{clarke-proximity}
	\bibinfo{author}{\bibfnamefont{J.}~\bibnamefont{Clark}}, \bibinfo{journal}{J.
		Phys. Colloq.} \textbf{\bibinfo{volume}{29}}, \bibinfo{pages}{C2}
	(\bibinfo{year}{1968}), ISSN \bibinfo{issn}{0449-1947}.
	
	\bibitem[{\citenamefont{Rai et~al.}(2019)\citenamefont{Rai, Haas, and
			Jagannathan}}]{PhysRevB.100.165121}
	\bibinfo{author}{\bibfnamefont{G.}~\bibnamefont{Rai}},
	\bibinfo{author}{\bibfnamefont{S.}~\bibnamefont{Haas}}, \bibnamefont{and}
	\bibinfo{author}{\bibfnamefont{A.}~\bibnamefont{Jagannathan}},
	\bibinfo{journal}{Phys. Rev. B} \textbf{\bibinfo{volume}{100}},
	\bibinfo{pages}{165121} (\bibinfo{year}{2019}),
	\urlprefix\url{https://link.aps.org/doi/10.1103/PhysRevB.100.165121}.
	
	\bibitem[{\citenamefont{Rai et~al.}(2020)\citenamefont{Rai, Haas, and
			Jagannathan}}]{PE-GRai}
	\bibinfo{author}{\bibfnamefont{G.}~\bibnamefont{Rai}},
	\bibinfo{author}{\bibfnamefont{S.}~\bibnamefont{Haas}}, \bibnamefont{and}
	\bibinfo{author}{\bibfnamefont{A.}~\bibnamefont{Jagannathan}},
	\bibinfo{journal}{Phys. Rev. B} \textbf{\bibinfo{volume}{102}},
	\bibinfo{pages}{134211} (\bibinfo{year}{2020}),
	\urlprefix\url{https://link.aps.org/doi/10.1103/PhysRevB.102.134211}.
	
	\bibitem[{\citenamefont{Kobes and Whitehead}(1987)}]{PE-RLkobes}
	\bibinfo{author}{\bibfnamefont{R.~L.} \bibnamefont{Kobes}} \bibnamefont{and}
	\bibinfo{author}{\bibfnamefont{J.~P.} \bibnamefont{Whitehead}},
	\bibinfo{journal}{Phys. Rev. B} \textbf{\bibinfo{volume}{36}},
	\bibinfo{pages}{121} (\bibinfo{year}{1987}),
	\urlprefix\url{https://link.aps.org/doi/10.1103/PhysRevB.36.121}.
	
	\bibitem[{\citenamefont{Chiu et~al.}(2016)\citenamefont{Chiu, Cole, and
			Das~Sarma}}]{PE-Ssharma}
	\bibinfo{author}{\bibfnamefont{C.-K.} \bibnamefont{Chiu}},
	\bibinfo{author}{\bibfnamefont{W.~S.} \bibnamefont{Cole}}, \bibnamefont{and}
	\bibinfo{author}{\bibfnamefont{S.}~\bibnamefont{Das~Sarma}},
	\bibinfo{journal}{Phys. Rev. B} \textbf{\bibinfo{volume}{94}},
	\bibinfo{pages}{125304} (\bibinfo{year}{2016}),
	\urlprefix\url{https://link.aps.org/doi/10.1103/PhysRevB.94.125304}.
	
	\bibitem[{\citenamefont{Yamazaki et~al.}(2006)\citenamefont{Yamazaki, Shannon,
			and Takagi}}]{PhysRevB.73.094507}
	\bibinfo{author}{\bibfnamefont{H.}~\bibnamefont{Yamazaki}},
	\bibinfo{author}{\bibfnamefont{N.}~\bibnamefont{Shannon}}, \bibnamefont{and}
	\bibinfo{author}{\bibfnamefont{H.}~\bibnamefont{Takagi}},
	\bibinfo{journal}{Phys. Rev. B} \textbf{\bibinfo{volume}{73}},
	\bibinfo{pages}{094507} (\bibinfo{year}{2006}),
	\urlprefix\url{https://link.aps.org/doi/10.1103/PhysRevB.73.094507}.
	
	\bibitem[{\citenamefont{Yamazaki et~al.}(2010)\citenamefont{Yamazaki, Shannon,
			and Takagi}}]{PhysRevB.81.094503}
	\bibinfo{author}{\bibfnamefont{H.}~\bibnamefont{Yamazaki}},
	\bibinfo{author}{\bibfnamefont{N.}~\bibnamefont{Shannon}}, \bibnamefont{and}
	\bibinfo{author}{\bibfnamefont{H.}~\bibnamefont{Takagi}},
	\bibinfo{journal}{Phys. Rev. B} \textbf{\bibinfo{volume}{81}},
	\bibinfo{pages}{094503} (\bibinfo{year}{2010}),
	\urlprefix\url{https://link.aps.org/doi/10.1103/PhysRevB.81.094503}.
	
	\bibitem[{\citenamefont{Csire et~al.}(2015)\citenamefont{Csire, \'Ujfalussy,
			Cserti, and Gy\ifmmode~\mbox{\H{o}}\else
			\H{o}\fi{}rffy}}]{PhysRevB.91.165142}
	\bibinfo{author}{\bibfnamefont{G.}~\bibnamefont{Csire}},
	\bibinfo{author}{\bibfnamefont{B.}~\bibnamefont{\'Ujfalussy}},
	\bibinfo{author}{\bibfnamefont{J.}~\bibnamefont{Cserti}}, \bibnamefont{and}
	\bibinfo{author}{\bibfnamefont{B.}~\bibnamefont{Gy\ifmmode~\mbox{\H{o}}\else
			\H{o}\fi{}rffy}}, \bibinfo{journal}{Phys. Rev. B}
	\textbf{\bibinfo{volume}{91}}, \bibinfo{pages}{165142}
	(\bibinfo{year}{2015}),
	\urlprefix\url{https://link.aps.org/doi/10.1103/PhysRevB.91.165142}.
	
	\bibitem[{\citenamefont{Csire et~al.}(2016{\natexlab{a}})\citenamefont{Csire,
			Cserti, T\"utt\ifmmode~\mbox{\H{o}}\else \H{o}\fi{}, and
			\'Ujfalussy}}]{PhysRevB.94.104511}
	\bibinfo{author}{\bibfnamefont{G.}~\bibnamefont{Csire}},
	\bibinfo{author}{\bibfnamefont{J.}~\bibnamefont{Cserti}},
	\bibinfo{author}{\bibfnamefont{I.}~\bibnamefont{T\"utt\ifmmode~\mbox{\H{o}}\else
			\H{o}\fi{}}}, \bibnamefont{and}
	\bibinfo{author}{\bibfnamefont{B.}~\bibnamefont{\'Ujfalussy}},
	\bibinfo{journal}{Phys. Rev. B} \textbf{\bibinfo{volume}{94}},
	\bibinfo{pages}{104511} (\bibinfo{year}{2016}{\natexlab{a}}),
	\urlprefix\url{https://link.aps.org/doi/10.1103/PhysRevB.94.104511}.
	
	\bibitem[{\citenamefont{Csire et~al.}(2016{\natexlab{b}})\citenamefont{Csire,
			Cserti, and Ujfalussy}}]{PE-GCsire}
	\bibinfo{author}{\bibfnamefont{G.}~\bibnamefont{Csire}},
	\bibinfo{author}{\bibfnamefont{J.}~\bibnamefont{Cserti}}, \bibnamefont{and}
	\bibinfo{author}{\bibfnamefont{B.}~\bibnamefont{Ujfalussy}},
	\bibinfo{journal}{J. Phys.: Condens. Matt.} \textbf{\bibinfo{volume}{28}},
	\bibinfo{pages}{495701} (\bibinfo{year}{2016}{\natexlab{b}}),
	\urlprefix\url{https://dx.doi.org/10.1088/0953-8984/28/49/495701}.
	
	\bibitem[{\citenamefont{Buzdin}(2005)}]{SC-FM-rev}
	\bibinfo{author}{\bibfnamefont{A.~I.} \bibnamefont{Buzdin}},
	\bibinfo{journal}{Rev. Mod. Phys.} \textbf{\bibinfo{volume}{77}},
	\bibinfo{pages}{935} (\bibinfo{year}{2005}),
	\urlprefix\url{https://link.aps.org/doi/10.1103/RevModPhys.77.935}.
	
	\bibitem[{\citenamefont{Fu and Kane}(2008)}]{PhysRevLett.100.096407}
	\bibinfo{author}{\bibfnamefont{L.}~\bibnamefont{Fu}} \bibnamefont{and}
	\bibinfo{author}{\bibfnamefont{C.~L.} \bibnamefont{Kane}},
	\bibinfo{journal}{Phys. Rev. Lett.} \textbf{\bibinfo{volume}{100}},
	\bibinfo{pages}{096407} (\bibinfo{year}{2008}),
	\urlprefix\url{https://link.aps.org/doi/10.1103/PhysRevLett.100.096407}.
	
	\bibitem[{\citenamefont{Kibble}(1976)}]{kibble-1976}
	\bibinfo{author}{\bibfnamefont{T.~W.~B.} \bibnamefont{Kibble}},
	\bibinfo{journal}{J. Phys. A: Math. Gen.} \textbf{\bibinfo{volume}{9}},
	\bibinfo{pages}{1387} (\bibinfo{year}{1976}),
	\urlprefix\url{https://dx.doi.org/10.1088/0305-4470/9/8/029}.
	
	\bibitem[{\citenamefont{Kibble}(1980)}]{kibble-1980}
	\bibinfo{author}{\bibfnamefont{T.}~\bibnamefont{Kibble}},
	\bibinfo{journal}{Phys. Rep.} \textbf{\bibinfo{volume}{67}},
	\bibinfo{pages}{183} (\bibinfo{year}{1980}), ISSN \bibinfo{issn}{0370-1573},
	\urlprefix\url{https://www.sciencedirect.com/science/article/pii/0370157380900915}.
	
	\bibitem[{\citenamefont{Zurek}(1985)}]{Zurek-1985}
	\bibinfo{author}{\bibfnamefont{W.~H.} \bibnamefont{Zurek}},
	\bibinfo{journal}{Nature (London)} \textbf{\bibinfo{volume}{317}},
	\bibinfo{pages}{505} (\bibinfo{year}{1985}), ISSN \bibinfo{issn}{0028-0836}.
	
	\bibitem[{\citenamefont{Zurek}(1996)}]{Zurek-1996}
	\bibinfo{author}{\bibfnamefont{W.}~\bibnamefont{Zurek}},
	\bibinfo{journal}{Phys. Rep.} \textbf{\bibinfo{volume}{276}},
	\bibinfo{pages}{177} (\bibinfo{year}{1996}),
	\urlprefix\url{https://doi.org/10.1016%2Fs0370-1573%2896%2900009-9}.
	
	\bibitem[{\citenamefont{Laguna and Zurek}(1997)}]{Zurek-1997}
	\bibinfo{author}{\bibfnamefont{P.}~\bibnamefont{Laguna}} \bibnamefont{and}
	\bibinfo{author}{\bibfnamefont{W.~H.} \bibnamefont{Zurek}},
	\bibinfo{journal}{Phys. Rev. Lett.} \textbf{\bibinfo{volume}{78}},
	\bibinfo{pages}{2519} (\bibinfo{year}{1997}),
	\urlprefix\url{https://link.aps.org/doi/10.1103/PhysRevLett.78.2519}.
	
	\bibitem[{\citenamefont{Laguna and Zurek}(1998)}]{Zurek-1998}
	\bibinfo{author}{\bibfnamefont{P.}~\bibnamefont{Laguna}} \bibnamefont{and}
	\bibinfo{author}{\bibfnamefont{W.~H.} \bibnamefont{Zurek}},
	\bibinfo{journal}{Phys. Rev. D} \textbf{\bibinfo{volume}{58}},
	\bibinfo{pages}{085021} (\bibinfo{year}{1998}),
	\urlprefix\url{https://link.aps.org/doi/10.1103/PhysRevD.58.085021}.
	
	\bibitem[{\citenamefont{Anglin and Zurek}(1999)}]{Zurek-1999}
	\bibinfo{author}{\bibfnamefont{J.~R.} \bibnamefont{Anglin}} \bibnamefont{and}
	\bibinfo{author}{\bibfnamefont{W.~H.} \bibnamefont{Zurek}},
	\bibinfo{journal}{Phys. Rev. Lett.} \textbf{\bibinfo{volume}{83}},
	\bibinfo{pages}{1707} (\bibinfo{year}{1999}),
	\urlprefix\url{https://link.aps.org/doi/10.1103/PhysRevLett.83.1707}.
	
	\bibitem[{\citenamefont{Stephens et~al.}(2002)\citenamefont{Stephens,
			Bettencourt, and Zurek}}]{Zurek-2002}
	\bibinfo{author}{\bibfnamefont{G.~J.} \bibnamefont{Stephens}},
	\bibinfo{author}{\bibfnamefont{L.~M.~A.} \bibnamefont{Bettencourt}},
	\bibnamefont{and} \bibinfo{author}{\bibfnamefont{W.~H.} \bibnamefont{Zurek}},
	\bibinfo{journal}{Phys. Rev. Lett.} \textbf{\bibinfo{volume}{88}},
	\bibinfo{pages}{137004} (\bibinfo{year}{2002}),
	\urlprefix\url{https://link.aps.org/doi/10.1103/PhysRevLett.88.137004}.
	
	\bibitem[{\citenamefont{Dziarmaga and Rams}(2010)}]{Dziarmaga-doiqpt}
	\bibinfo{author}{\bibfnamefont{J.}~\bibnamefont{Dziarmaga}} \bibnamefont{and}
	\bibinfo{author}{\bibfnamefont{M.~M.} \bibnamefont{Rams}},
	\bibinfo{journal}{New J. Phys.} \textbf{\bibinfo{volume}{12}},
	\bibinfo{pages}{055007} (\bibinfo{year}{2010}), ISSN
	\bibinfo{issn}{1367-2630}.
	
	\bibitem[{\citenamefont{Dziarmaga}(2005)}]{QPT-Dziarmaga-2005}
	\bibinfo{author}{\bibfnamefont{J.}~\bibnamefont{Dziarmaga}},
	\bibinfo{journal}{Phys. Rev. Lett.} \textbf{\bibinfo{volume}{95}},
	\bibinfo{pages}{245701} (\bibinfo{year}{2005}),
	\urlprefix\url{https://link.aps.org/doi/10.1103/PhysRevLett.95.245701}.
	
	\bibitem[{\citenamefont{Uhlmann
			et~al.}(2010{\natexlab{a}})\citenamefont{Uhlmann, Sch\"utzhold, and
			Fischer}}]{PhysRevD.81.025017}
	\bibinfo{author}{\bibfnamefont{M.}~\bibnamefont{Uhlmann}},
	\bibinfo{author}{\bibfnamefont{R.}~\bibnamefont{Sch\"utzhold}},
	\bibnamefont{and} \bibinfo{author}{\bibfnamefont{U.~R.}
		\bibnamefont{Fischer}}, \bibinfo{journal}{Phys. Rev. D}
	\textbf{\bibinfo{volume}{81}}, \bibinfo{pages}{025017}
	(\bibinfo{year}{2010}{\natexlab{a}}),
	\urlprefix\url{https://link.aps.org/doi/10.1103/PhysRevD.81.025017}.
	
	\bibitem[{\citenamefont{Uhlmann
			et~al.}(2010{\natexlab{b}})\citenamefont{Uhlmann, Schützhold, and
			Fischer}}]{Uhlmann_2010}
	\bibinfo{author}{\bibfnamefont{M.}~\bibnamefont{Uhlmann}},
	\bibinfo{author}{\bibfnamefont{R.}~\bibnamefont{Schützhold}},
	\bibnamefont{and} \bibinfo{author}{\bibfnamefont{U.~R.}
		\bibnamefont{Fischer}}, \bibinfo{journal}{New J. Phys.}
	\textbf{\bibinfo{volume}{12}}, \bibinfo{pages}{095020}
	(\bibinfo{year}{2010}{\natexlab{b}}),
	\urlprefix\url{https://dx.doi.org/10.1088/1367-2630/12/9/095020}.
	
	\bibitem[{\citenamefont{Dziarmaga et~al.}(2022)\citenamefont{Dziarmaga, Rams,
			and Zurek}}]{Dziarmaga-2022}
	\bibinfo{author}{\bibfnamefont{J.}~\bibnamefont{Dziarmaga}},
	\bibinfo{author}{\bibfnamefont{M.~M.} \bibnamefont{Rams}}, \bibnamefont{and}
	\bibinfo{author}{\bibfnamefont{W.~H.} \bibnamefont{Zurek}},
	\bibinfo{journal}{Phys. Rev. Lett.} \textbf{\bibinfo{volume}{129}},
	\bibinfo{pages}{260407} (\bibinfo{year}{2022}),
	\urlprefix\url{https://doi.org/10.1103/physrevlett.129.260407}.
	
	\bibitem[{\citenamefont{Jaschke et~al.}(2017)\citenamefont{Jaschke, Maeda,
			Whalen, Wall, and Carr}}]{QIM-Jaschke-2017}
	\bibinfo{author}{\bibfnamefont{D.}~\bibnamefont{Jaschke}},
	\bibinfo{author}{\bibfnamefont{K.}~\bibnamefont{Maeda}},
	\bibinfo{author}{\bibfnamefont{J.~D.} \bibnamefont{Whalen}},
	\bibinfo{author}{\bibfnamefont{M.~L.} \bibnamefont{Wall}}, \bibnamefont{and}
	\bibinfo{author}{\bibfnamefont{L.~D.} \bibnamefont{Carr}},
	\bibinfo{journal}{New J. Phys.} \textbf{\bibinfo{volume}{19}},
	\bibinfo{pages}{033032} (\bibinfo{year}{2017}),
	\urlprefix\url{https://dx.doi.org/10.1088/1367-2630/aa65bc}.
	
	\bibitem[{\citenamefont{Polkovnikov}(2005)}]{adiabtic-dynamics-2005}
	\bibinfo{author}{\bibfnamefont{A.}~\bibnamefont{Polkovnikov}},
	\bibinfo{journal}{Phys. Rev. B} \textbf{\bibinfo{volume}{72}},
	\bibinfo{pages}{161201(R)} (\bibinfo{year}{2005}),
	\urlprefix\url{https://link.aps.org/doi/10.1103/PhysRevB.72.161201}.
	
	\bibitem[{\citenamefont{Damski}(2005)}]{Damski-2005}
	\bibinfo{author}{\bibfnamefont{B.}~\bibnamefont{Damski}},
	\bibinfo{journal}{Phys. Rev. Lett.} \textbf{\bibinfo{volume}{95}},
	\bibinfo{pages}{035701} (\bibinfo{year}{2005}),
	\urlprefix\url{https://link.aps.org/doi/10.1103/PhysRevLett.95.035701}.
	
	\bibitem[{\citenamefont{Warner and Leggett}(2005)}]{QD-Legget-2005}
	\bibinfo{author}{\bibfnamefont{G.~L.} \bibnamefont{Warner}} \bibnamefont{and}
	\bibinfo{author}{\bibfnamefont{A.~J.} \bibnamefont{Leggett}},
	\bibinfo{journal}{Phys. Rev. B} \textbf{\bibinfo{volume}{71}},
	\bibinfo{pages}{134514} (\bibinfo{year}{2005}),
	\urlprefix\url{https://link.aps.org/doi/10.1103/PhysRevB.71.134514}.
	
	\bibitem[{\citenamefont{Zurek et~al.}(2005)\citenamefont{Zurek, Dorner, and
			Zoller}}]{QPT-Zurek-2005}
	\bibinfo{author}{\bibfnamefont{W.~H.} \bibnamefont{Zurek}},
	\bibinfo{author}{\bibfnamefont{U.}~\bibnamefont{Dorner}}, \bibnamefont{and}
	\bibinfo{author}{\bibfnamefont{P.}~\bibnamefont{Zoller}},
	\bibinfo{journal}{Phys. Rev. Lett.} \textbf{\bibinfo{volume}{95}},
	\bibinfo{pages}{105701} (\bibinfo{year}{2005}),
	\urlprefix\url{https://link.aps.org/doi/10.1103/PhysRevLett.95.105701}.
	
	\bibitem[{\citenamefont{Zurek and Dorner}(2008)}]{Zurekptis}
	\bibinfo{author}{\bibfnamefont{W.~H.} \bibnamefont{Zurek}} \bibnamefont{and}
	\bibinfo{author}{\bibfnamefont{U.}~\bibnamefont{Dorner}},
	\bibinfo{journal}{Philos. Trans. Royal Soc. A}
	\textbf{\bibinfo{volume}{366}}, \bibinfo{pages}{2953} (\bibinfo{year}{2008}),
	ISSN \bibinfo{issn}{1364-503X}.
	
	\bibitem[{\citenamefont{Shimizu et~al.}(2018)\citenamefont{Shimizu, Kuno,
			Hirano, and Ichinose}}]{PhysRevA.97.033626}
	\bibinfo{author}{\bibfnamefont{K.}~\bibnamefont{Shimizu}},
	\bibinfo{author}{\bibfnamefont{Y.}~\bibnamefont{Kuno}},
	\bibinfo{author}{\bibfnamefont{T.}~\bibnamefont{Hirano}}, \bibnamefont{and}
	\bibinfo{author}{\bibfnamefont{I.}~\bibnamefont{Ichinose}},
	\bibinfo{journal}{Phys. Rev. A} \textbf{\bibinfo{volume}{97}},
	\bibinfo{pages}{033626} (\bibinfo{year}{2018}),
	\urlprefix\url{https://link.aps.org/doi/10.1103/PhysRevA.97.033626}.
	
	\bibitem[{\citenamefont{Cucchietti et~al.}(2007)\citenamefont{Cucchietti,
			Damski, Dziarmaga, and Zurek}}]{PhysRevA.75.023603}
	\bibinfo{author}{\bibfnamefont{F.~M.} \bibnamefont{Cucchietti}},
	\bibinfo{author}{\bibfnamefont{B.}~\bibnamefont{Damski}},
	\bibinfo{author}{\bibfnamefont{J.}~\bibnamefont{Dziarmaga}},
	\bibnamefont{and} \bibinfo{author}{\bibfnamefont{W.~H.} \bibnamefont{Zurek}},
	\bibinfo{journal}{Phys. Rev. A} \textbf{\bibinfo{volume}{75}},
	\bibinfo{pages}{023603} (\bibinfo{year}{2007}),
	\urlprefix\url{https://link.aps.org/doi/10.1103/PhysRevA.75.023603}.
	
	\bibitem[{\citenamefont{Dziarmaga et~al.}(2012)\citenamefont{Dziarmaga,
			Tylutki, and Zurek}}]{PhysRevB.86.144521}
	\bibinfo{author}{\bibfnamefont{J.}~\bibnamefont{Dziarmaga}},
	\bibinfo{author}{\bibfnamefont{M.}~\bibnamefont{Tylutki}}, \bibnamefont{and}
	\bibinfo{author}{\bibfnamefont{W.~H.} \bibnamefont{Zurek}},
	\bibinfo{journal}{Phys. Rev. B} \textbf{\bibinfo{volume}{86}},
	\bibinfo{pages}{144521} (\bibinfo{year}{2012}),
	\urlprefix\url{https://link.aps.org/doi/10.1103/PhysRevB.86.144521}.
	
	\bibitem[{\citenamefont{Gardas et~al.}(2017)\citenamefont{Gardas, Dziarmaga,
			and Zurek}}]{PhysRevB.95.104306}
	\bibinfo{author}{\bibfnamefont{B.}~\bibnamefont{Gardas}},
	\bibinfo{author}{\bibfnamefont{J.}~\bibnamefont{Dziarmaga}},
	\bibnamefont{and} \bibinfo{author}{\bibfnamefont{W.~H.} \bibnamefont{Zurek}},
	\bibinfo{journal}{Phys. Rev. B} \textbf{\bibinfo{volume}{95}},
	\bibinfo{pages}{104306} (\bibinfo{year}{2017}),
	\urlprefix\url{https://link.aps.org/doi/10.1103/PhysRevB.95.104306}.
	
	\bibitem[{\citenamefont{Machida and Kasamatsu}(2021)}]{PhysRevA.103.013310}
	\bibinfo{author}{\bibfnamefont{Y.}~\bibnamefont{Machida}} \bibnamefont{and}
	\bibinfo{author}{\bibfnamefont{K.}~\bibnamefont{Kasamatsu}},
	\bibinfo{journal}{Phys. Rev. A} \textbf{\bibinfo{volume}{103}},
	\bibinfo{pages}{013310} (\bibinfo{year}{2021}),
	\urlprefix\url{https://link.aps.org/doi/10.1103/PhysRevA.103.013310}.
	
	\bibitem[{\citenamefont{Anquez et~al.}(2016{\natexlab{a}})\citenamefont{Anquez,
			Robbins, Bharath, Boguslawski, Hoang, and Chapman}}]{PhysRevLett.116.155301}
	\bibinfo{author}{\bibfnamefont{M.}~\bibnamefont{Anquez}},
	\bibinfo{author}{\bibfnamefont{B.~A.} \bibnamefont{Robbins}},
	\bibinfo{author}{\bibfnamefont{H.~M.} \bibnamefont{Bharath}},
	\bibinfo{author}{\bibfnamefont{M.}~\bibnamefont{Boguslawski}},
	\bibinfo{author}{\bibfnamefont{T.~M.} \bibnamefont{Hoang}}, \bibnamefont{and}
	\bibinfo{author}{\bibfnamefont{M.~S.} \bibnamefont{Chapman}},
	\bibinfo{journal}{Phys. Rev. Lett.} \textbf{\bibinfo{volume}{116}},
	\bibinfo{pages}{155301} (\bibinfo{year}{2016}{\natexlab{a}}),
	\urlprefix\url{https://link.aps.org/doi/10.1103/PhysRevLett.116.155301}.
	
	\bibitem[{\citenamefont{Damski and Zurek}(2009)}]{Damski_2009}
	\bibinfo{author}{\bibfnamefont{B.}~\bibnamefont{Damski}} \bibnamefont{and}
	\bibinfo{author}{\bibfnamefont{W.~H.} \bibnamefont{Zurek}},
	\bibinfo{journal}{New J. Phys.} \textbf{\bibinfo{volume}{11}},
	\bibinfo{pages}{063014} (\bibinfo{year}{2009}),
	\urlprefix\url{https://dx.doi.org/10.1088/1367-2630/11/6/063014}.
	
	\bibitem[{\citenamefont{Monaco et~al.}(2002)\citenamefont{Monaco, Mygind, and
			Rivers}}]{Monaco-2002}
	\bibinfo{author}{\bibfnamefont{R.}~\bibnamefont{Monaco}},
	\bibinfo{author}{\bibfnamefont{J.}~\bibnamefont{Mygind}}, \bibnamefont{and}
	\bibinfo{author}{\bibfnamefont{R.~J.} \bibnamefont{Rivers}},
	\bibinfo{journal}{Phys. Rev. Lett.} \textbf{\bibinfo{volume}{89}},
	\bibinfo{pages}{080603} (\bibinfo{year}{2002}),
	\urlprefix\url{https://link.aps.org/doi/10.1103/PhysRevLett.89.080603}.
	
	\bibitem[{\citenamefont{Ulm et~al.}(2013)\citenamefont{Ulm, Rossnagel, Jacob,
			Deguenther, Dawkins, Poschinger, Nigmatullin, Retzker, Plenio, Schmidt-Kaler
			et~al.}}]{Ulm-2013}
	\bibinfo{author}{\bibfnamefont{S.}~\bibnamefont{Ulm}},
	\bibinfo{author}{\bibfnamefont{J.}~\bibnamefont{Rossnagel}},
	\bibinfo{author}{\bibfnamefont{G.}~\bibnamefont{Jacob}},
	\bibinfo{author}{\bibfnamefont{C.}~\bibnamefont{Deguenther}},
	\bibinfo{author}{\bibfnamefont{S.~T.} \bibnamefont{Dawkins}},
	\bibinfo{author}{\bibfnamefont{U.~G.} \bibnamefont{Poschinger}},
	\bibinfo{author}{\bibfnamefont{R.}~\bibnamefont{Nigmatullin}},
	\bibinfo{author}{\bibfnamefont{A.}~\bibnamefont{Retzker}},
	\bibinfo{author}{\bibfnamefont{M.~B.} \bibnamefont{Plenio}},
	\bibinfo{author}{\bibfnamefont{F.}~\bibnamefont{Schmidt-Kaler}},
	\bibnamefont{et~al.}, \bibinfo{journal}{Nat. Commun.}
	\textbf{\bibinfo{volume}{4}}, \bibinfo{pages}{2290} (\bibinfo{year}{2013}),
	ISSN \bibinfo{issn}{2041-1723}.
	
	\bibitem[{\citenamefont{Pyka et~al.}(2013)\citenamefont{Pyka, Keller, Partner,
			Nigmatullin, Burgermeister, Meier, Kuhlmann, Retzker, Plenio, Zurek
			et~al.}}]{Pyka-2013}
	\bibinfo{author}{\bibfnamefont{K.}~\bibnamefont{Pyka}},
	\bibinfo{author}{\bibfnamefont{J.}~\bibnamefont{Keller}},
	\bibinfo{author}{\bibfnamefont{H.~L.} \bibnamefont{Partner}},
	\bibinfo{author}{\bibfnamefont{R.}~\bibnamefont{Nigmatullin}},
	\bibinfo{author}{\bibfnamefont{T.}~\bibnamefont{Burgermeister}},
	\bibinfo{author}{\bibfnamefont{D.~M.} \bibnamefont{Meier}},
	\bibinfo{author}{\bibfnamefont{K.}~\bibnamefont{Kuhlmann}},
	\bibinfo{author}{\bibfnamefont{A.}~\bibnamefont{Retzker}},
	\bibinfo{author}{\bibfnamefont{M.~B.} \bibnamefont{Plenio}},
	\bibinfo{author}{\bibfnamefont{W.~H.} \bibnamefont{Zurek}},
	\bibnamefont{et~al.}, \bibinfo{journal}{Nat. Commun.}
	\textbf{\bibinfo{volume}{4}}, \bibinfo{pages}{2291} (\bibinfo{year}{2013}),
	ISSN \bibinfo{issn}{2041-1723}.
	
	\bibitem[{\citenamefont{Navon et~al.}(2015)\citenamefont{Navon, Gaunt, Smith,
			and Hadzibabic}}]{ALGaunt-2015}
	\bibinfo{author}{\bibfnamefont{N.}~\bibnamefont{Navon}},
	\bibinfo{author}{\bibfnamefont{A.~L.} \bibnamefont{Gaunt}},
	\bibinfo{author}{\bibfnamefont{R.~P.} \bibnamefont{Smith}}, \bibnamefont{and}
	\bibinfo{author}{\bibfnamefont{Z.}~\bibnamefont{Hadzibabic}},
	\bibinfo{journal}{Science} \textbf{\bibinfo{volume}{347}},
	\bibinfo{pages}{167} (\bibinfo{year}{2015}),
	\eprint{https://www.science.org/doi/pdf/10.1126/science.1258676},
	\urlprefix\url{https://www.science.org/doi/abs/10.1126/science.1258676}.
	
	\bibitem[{\citenamefont{Braun et~al.}(2015)\citenamefont{Braun, Friesdorf,
			Hodgman, Schreiber, Ronzheimer, Riera, del Rey, Bloch, Eisert, and
			Schneider}}]{QPT-Brawn-2015}
	\bibinfo{author}{\bibfnamefont{S.}~\bibnamefont{Braun}},
	\bibinfo{author}{\bibfnamefont{M.}~\bibnamefont{Friesdorf}},
	\bibinfo{author}{\bibfnamefont{S.~S.} \bibnamefont{Hodgman}},
	\bibinfo{author}{\bibfnamefont{M.}~\bibnamefont{Schreiber}},
	\bibinfo{author}{\bibfnamefont{J.~P.} \bibnamefont{Ronzheimer}},
	\bibinfo{author}{\bibfnamefont{A.}~\bibnamefont{Riera}},
	\bibinfo{author}{\bibfnamefont{M.}~\bibnamefont{del Rey}},
	\bibinfo{author}{\bibfnamefont{I.}~\bibnamefont{Bloch}},
	\bibinfo{author}{\bibfnamefont{J.}~\bibnamefont{Eisert}}, \bibnamefont{and}
	\bibinfo{author}{\bibfnamefont{U.}~\bibnamefont{Schneider}},
	\bibinfo{journal}{PNAS} \textbf{\bibinfo{volume}{112}}, \bibinfo{pages}{3641}
	(\bibinfo{year}{2015}),
	\eprint{https://www.pnas.org/doi/pdf/10.1073/pnas.1408861112},
	\urlprefix\url{https://www.pnas.org/doi/abs/10.1073/pnas.1408861112}.
	
	\bibitem[{\citenamefont{Chen et~al.}(2011)\citenamefont{Chen, White, Borries,
			and DeMarco}}]{QPT-mott-2011}
	\bibinfo{author}{\bibfnamefont{D.}~\bibnamefont{Chen}},
	\bibinfo{author}{\bibfnamefont{M.}~\bibnamefont{White}},
	\bibinfo{author}{\bibfnamefont{C.}~\bibnamefont{Borries}}, \bibnamefont{and}
	\bibinfo{author}{\bibfnamefont{B.}~\bibnamefont{DeMarco}},
	\bibinfo{journal}{Phys. Rev. Lett.} \textbf{\bibinfo{volume}{106}},
	\bibinfo{pages}{235304} (\bibinfo{year}{2011}),
	\urlprefix\url{https://link.aps.org/doi/10.1103/PhysRevLett.106.235304}.
	
	\bibitem[{\citenamefont{Keesling et~al.}(2019)\citenamefont{Keesling, Omran,
			Levine, Bernien, Pichler, Choi, Samajdar, Schwartz, Silvi, Sachdev
			et~al.}}]{QKZM-Nature-2019}
	\bibinfo{author}{\bibfnamefont{A.}~\bibnamefont{Keesling}},
	\bibinfo{author}{\bibfnamefont{A.}~\bibnamefont{Omran}},
	\bibinfo{author}{\bibfnamefont{H.}~\bibnamefont{Levine}},
	\bibinfo{author}{\bibfnamefont{H.}~\bibnamefont{Bernien}},
	\bibinfo{author}{\bibfnamefont{H.}~\bibnamefont{Pichler}},
	\bibinfo{author}{\bibfnamefont{S.}~\bibnamefont{Choi}},
	\bibinfo{author}{\bibfnamefont{R.}~\bibnamefont{Samajdar}},
	\bibinfo{author}{\bibfnamefont{S.}~\bibnamefont{Schwartz}},
	\bibinfo{author}{\bibfnamefont{P.}~\bibnamefont{Silvi}},
	\bibinfo{author}{\bibfnamefont{S.}~\bibnamefont{Sachdev}},
	\bibnamefont{et~al.}, \bibinfo{journal}{Nature}
	\textbf{\bibinfo{volume}{568}}, \bibinfo{pages}{207} (\bibinfo{year}{2019}),
	ISSN \bibinfo{issn}{0028-0836}.
	
	\bibitem[{\citenamefont{Anquez et~al.}(2016{\natexlab{b}})\citenamefont{Anquez,
			Robbins, Bharath, Boguslawski, Hoang, and Chapman}}]{QKZM-prl-2016}
	\bibinfo{author}{\bibfnamefont{M.}~\bibnamefont{Anquez}},
	\bibinfo{author}{\bibfnamefont{B.~A.} \bibnamefont{Robbins}},
	\bibinfo{author}{\bibfnamefont{H.~M.} \bibnamefont{Bharath}},
	\bibinfo{author}{\bibfnamefont{M.}~\bibnamefont{Boguslawski}},
	\bibinfo{author}{\bibfnamefont{T.~M.} \bibnamefont{Hoang}}, \bibnamefont{and}
	\bibinfo{author}{\bibfnamefont{M.~S.} \bibnamefont{Chapman}},
	\bibinfo{journal}{Phys. Rev. Lett.} \textbf{\bibinfo{volume}{116}},
	\bibinfo{pages}{155301} (\bibinfo{year}{2016}{\natexlab{b}}),
	\urlprefix\url{https://link.aps.org/doi/10.1103/PhysRevLett.116.155301}.
	
	\bibitem[{\citenamefont{Li et~al.}(2022)\citenamefont{Li, Wu, Mei, Yao, Lian,
			Cai, Wang, Qi, Yao, He et~al.}}]{TFIM-BWLi-2022}
	\bibinfo{author}{\bibfnamefont{B.~W.} \bibnamefont{Li}},
	\bibinfo{author}{\bibfnamefont{Y.~K.} \bibnamefont{Wu}},
	\bibinfo{author}{\bibfnamefont{Q.~X.} \bibnamefont{Mei}},
	\bibinfo{author}{\bibfnamefont{R.}~\bibnamefont{Yao}},
	\bibinfo{author}{\bibfnamefont{W.~Q.} \bibnamefont{Lian}},
	\bibinfo{author}{\bibfnamefont{M.~L.} \bibnamefont{Cai}},
	\bibinfo{author}{\bibfnamefont{Y.}~\bibnamefont{Wang}},
	\bibinfo{author}{\bibfnamefont{B.~X.} \bibnamefont{Qi}},
	\bibinfo{author}{\bibfnamefont{L.}~\bibnamefont{Yao}},
	\bibinfo{author}{\bibfnamefont{L.}~\bibnamefont{He}}, \bibnamefont{et~al.},
	\emph{\bibinfo{title}{Probing critical behavior of long-range
			transverse-field ising model through quantum kibble-zurek mechanism}}
	(\bibinfo{year}{2022}), \bibinfo{note}{arXiv:2208.03060},
	\urlprefix\url{https://arxiv.org/abs/2208.03060}.
	
	\bibitem[{\citenamefont{Deutschländer
			et~al.}(2015)\citenamefont{Deutschländer, Dillmann, Maret, and
			Keim}}]{KZM-colloidal-monolayer-2015}
	\bibinfo{author}{\bibfnamefont{S.}~\bibnamefont{Deutschländer}},
	\bibinfo{author}{\bibfnamefont{P.}~\bibnamefont{Dillmann}},
	\bibinfo{author}{\bibfnamefont{G.}~\bibnamefont{Maret}}, \bibnamefont{and}
	\bibinfo{author}{\bibfnamefont{P.}~\bibnamefont{Keim}},
	\bibinfo{journal}{PNAS} \textbf{\bibinfo{volume}{112}}, \bibinfo{pages}{6925}
	(\bibinfo{year}{2015}),
	\eprint{https://www.pnas.org/doi/pdf/10.1073/pnas.1500763112},
	\urlprefix\url{https://www.pnas.org/doi/abs/10.1073/pnas.1500763112}.
	
	\bibitem[{\citenamefont{Ko et~al.}(2019)\citenamefont{Ko, Park, and
			Shin}}]{KZM-FermiSF-2019}
	\bibinfo{author}{\bibfnamefont{B.}~\bibnamefont{Ko}},
	\bibinfo{author}{\bibfnamefont{J.~W.} \bibnamefont{Park}}, \bibnamefont{and}
	\bibinfo{author}{\bibfnamefont{Y.}~\bibnamefont{Shin}},
	\bibinfo{journal}{Nat. Phys.} \textbf{\bibinfo{volume}{15}},
	\bibinfo{pages}{1227} (\bibinfo{year}{2019}), ISSN \bibinfo{issn}{1745-2473}.
	
	\bibitem[{\citenamefont{Liu et~al.}(2021)\citenamefont{Liu, Yao, Deng, Wang,
			Wang, Li, Chen, Chen, and Pan}}]{KZM-vortices-2021}
	\bibinfo{author}{\bibfnamefont{X.-P.} \bibnamefont{Liu}},
	\bibinfo{author}{\bibfnamefont{X.-C.} \bibnamefont{Yao}},
	\bibinfo{author}{\bibfnamefont{Y.}~\bibnamefont{Deng}},
	\bibinfo{author}{\bibfnamefont{Y.-X.} \bibnamefont{Wang}},
	\bibinfo{author}{\bibfnamefont{X.-Q.} \bibnamefont{Wang}},
	\bibinfo{author}{\bibfnamefont{X.}~\bibnamefont{Li}},
	\bibinfo{author}{\bibfnamefont{Q.}~\bibnamefont{Chen}},
	\bibinfo{author}{\bibfnamefont{Y.-A.} \bibnamefont{Chen}}, \bibnamefont{and}
	\bibinfo{author}{\bibfnamefont{J.-W.} \bibnamefont{Pan}},
	\bibinfo{journal}{Phys. Rev. Res.} \textbf{\bibinfo{volume}{3}},
	\bibinfo{pages}{043115} (\bibinfo{year}{2021}),
	\urlprefix\url{https://link.aps.org/doi/10.1103/PhysRevResearch.3.043115}.
	
	\bibitem[{\citenamefont{Uhlmann et~al.}(2007)\citenamefont{Uhlmann,
			Sch\"utzhold, and Fischer}}]{PhysRevLett.99.120407}
	\bibinfo{author}{\bibfnamefont{M.}~\bibnamefont{Uhlmann}},
	\bibinfo{author}{\bibfnamefont{R.}~\bibnamefont{Sch\"utzhold}},
	\bibnamefont{and} \bibinfo{author}{\bibfnamefont{U.~R.}
		\bibnamefont{Fischer}}, \bibinfo{journal}{Phys. Rev. Lett.}
	\textbf{\bibinfo{volume}{99}}, \bibinfo{pages}{120407}
	(\bibinfo{year}{2007}),
	\urlprefix\url{https://link.aps.org/doi/10.1103/PhysRevLett.99.120407}.
	
	\bibitem[{\citenamefont{Chien and Damski}(2010)}]{Chien10}
	\bibinfo{author}{\bibfnamefont{C.~C.} \bibnamefont{Chien}} \bibnamefont{and}
	\bibinfo{author}{\bibfnamefont{B.}~\bibnamefont{Damski}},
	\bibinfo{journal}{Phys. Rev. A} \textbf{\bibinfo{volume}{82}},
	\bibinfo{pages}{063616} (\bibinfo{year}{2010}),
	\urlprefix\url{https://doi.org/10.1103/PhysRevA.82.063616}.
	
	\bibitem[{\citenamefont{G\'omez-Ruiz and del
			Campo}(2019)}]{PhysRevLett.122.080604}
	\bibinfo{author}{\bibfnamefont{F.~J.} \bibnamefont{G\'omez-Ruiz}}
	\bibnamefont{and} \bibinfo{author}{\bibfnamefont{A.}~\bibnamefont{del
			Campo}}, \bibinfo{journal}{Phys. Rev. Lett.} \textbf{\bibinfo{volume}{122}},
	\bibinfo{pages}{080604} (\bibinfo{year}{2019}),
	\urlprefix\url{https://link.aps.org/doi/10.1103/PhysRevLett.122.080604}.
	
	\bibitem[{\citenamefont{Bando et~al.}(2020)\citenamefont{Bando, Susa, Oshiyama,
			Shibata, Ohzeki, G\'omez-Ruiz, Lidar, Suzuki, del Campo, and
			Nishimori}}]{PhysRevResearch.2.033369}
	\bibinfo{author}{\bibfnamefont{Y.}~\bibnamefont{Bando}},
	\bibinfo{author}{\bibfnamefont{Y.}~\bibnamefont{Susa}},
	\bibinfo{author}{\bibfnamefont{H.}~\bibnamefont{Oshiyama}},
	\bibinfo{author}{\bibfnamefont{N.}~\bibnamefont{Shibata}},
	\bibinfo{author}{\bibfnamefont{M.}~\bibnamefont{Ohzeki}},
	\bibinfo{author}{\bibfnamefont{F.~J.} \bibnamefont{G\'omez-Ruiz}},
	\bibinfo{author}{\bibfnamefont{D.~A.} \bibnamefont{Lidar}},
	\bibinfo{author}{\bibfnamefont{S.}~\bibnamefont{Suzuki}},
	\bibinfo{author}{\bibfnamefont{A.}~\bibnamefont{del Campo}},
	\bibnamefont{and}
	\bibinfo{author}{\bibfnamefont{H.}~\bibnamefont{Nishimori}},
	\bibinfo{journal}{Phys. Rev. Res.} \textbf{\bibinfo{volume}{2}},
	\bibinfo{pages}{033369} (\bibinfo{year}{2020}),
	\urlprefix\url{https://link.aps.org/doi/10.1103/PhysRevResearch.2.033369}.
	
	\bibitem[{\citenamefont{G\'omez-Ruiz et~al.}(2020)\citenamefont{G\'omez-Ruiz,
			Mayo, and del Campo}}]{PhysRevLett.124.240602}
	\bibinfo{author}{\bibfnamefont{F.~J.} \bibnamefont{G\'omez-Ruiz}},
	\bibinfo{author}{\bibfnamefont{J.~J.} \bibnamefont{Mayo}}, \bibnamefont{and}
	\bibinfo{author}{\bibfnamefont{A.}~\bibnamefont{del Campo}},
	\bibinfo{journal}{Phys. Rev. Lett.} \textbf{\bibinfo{volume}{124}},
	\bibinfo{pages}{240602} (\bibinfo{year}{2020}),
	\urlprefix\url{https://link.aps.org/doi/10.1103/PhysRevLett.124.240602}.
	
	\bibitem[{\citenamefont{Cui et~al.}(2020)\citenamefont{Cui, G{\'o}mez-Ruiz,
			Huang, Li, Guo, and del Campo}}]{Commun.Phys.3.44}
	\bibinfo{author}{\bibfnamefont{J.-M.} \bibnamefont{Cui}},
	\bibinfo{author}{\bibfnamefont{F.~J.} \bibnamefont{G{\'o}mez-Ruiz}},
	\bibinfo{author}{\bibfnamefont{Y.-F.} \bibnamefont{Huang}},
	\bibinfo{author}{\bibfnamefont{C.-F.} \bibnamefont{Li}},
	\bibinfo{author}{\bibfnamefont{G.-C.} \bibnamefont{Guo}}, \bibnamefont{and}
	\bibinfo{author}{\bibfnamefont{A.}~\bibnamefont{del Campo}},
	\bibinfo{journal}{Communications Physics} \textbf{\bibinfo{volume}{3}},
	\bibinfo{pages}{44} (\bibinfo{year}{2020}),
	\urlprefix\url{https://doi.org/10.1038/s42005-020-0306-6}.
	
	\bibitem[{\citenamefont{Chaikin and Lubensky}(1995)}]{ChaikinP.M1995PoCM}
	\bibinfo{author}{\bibfnamefont{P.~M.} \bibnamefont{Chaikin}} \bibnamefont{and}
	\bibinfo{author}{\bibfnamefont{T.~C.} \bibnamefont{Lubensky}},
	\emph{\bibinfo{title}{Principles of Condensed Matter Physics}}
	(\bibinfo{publisher}{Cambridge University Press},
	\bibinfo{address}{Cambridge, UK}, \bibinfo{year}{1995}), ISBN
	\bibinfo{isbn}{9780521432245}.
	
	\bibitem[{\citenamefont{Chien}(2012)}]{Chien12}
	\bibinfo{author}{\bibfnamefont{C.~C.} \bibnamefont{Chien}},
	\bibinfo{journal}{Phys. Lett. A} \textbf{\bibinfo{volume}{376}},
	\bibinfo{pages}{729} (\bibinfo{year}{2012}),
	\urlprefix\url{https://doi.org/10.1016/j.physleta.2011.11.037}.
	
	\bibitem[{\citenamefont{Piselli et~al.}(2018)\citenamefont{Piselli, Simonucci,
			and Strinati}}]{PhysRevB.98.144508}
	\bibinfo{author}{\bibfnamefont{V.}~\bibnamefont{Piselli}},
	\bibinfo{author}{\bibfnamefont{S.}~\bibnamefont{Simonucci}},
	\bibnamefont{and} \bibinfo{author}{\bibfnamefont{G.~C.}
		\bibnamefont{Strinati}}, \bibinfo{journal}{Phys. Rev. B}
	\textbf{\bibinfo{volume}{98}}, \bibinfo{pages}{144508}
	(\bibinfo{year}{2018}),
	\urlprefix\url{https://link.aps.org/doi/10.1103/PhysRevB.98.144508}.
	
	\bibitem[{\citenamefont{Bogoliubov}(1947)}]{bogoliubov1947theory}
	\bibinfo{author}{\bibfnamefont{N.}~\bibnamefont{Bogoliubov}},
	\bibinfo{journal}{J. Phys.} \textbf{\bibinfo{volume}{11}},
	\bibinfo{pages}{23} (\bibinfo{year}{1947}).
	
	\bibitem[{\citenamefont{Zhu}(2016)}]{BdG-book}
	\bibinfo{author}{\bibfnamefont{J.-X.} \bibnamefont{Zhu}},
	\emph{\bibinfo{title}{Bogoliubov-de Gennes Method and Its Applications}},
	Lecture Notes in Physics, 924 (\bibinfo{publisher}{Springer International
		Publishing}, \bibinfo{address}{Cham}, \bibinfo{year}{2016}),
	\bibinfo{edition}{1st} ed., ISBN \bibinfo{isbn}{3-319-31314-2}.
	
	\bibitem[{\citenamefont{Fetter and Walecka}(1971)}]{Fetter_book}
	\bibinfo{author}{\bibfnamefont{A.~L.} \bibnamefont{Fetter}} \bibnamefont{and}
	\bibinfo{author}{\bibfnamefont{J.~D.} \bibnamefont{Walecka}},
	\emph{\bibinfo{title}{Quantum Theory of Many-Particle Systems}}
	(\bibinfo{publisher}{McGraw-Hill}, \bibinfo{address}{Boston},
	\bibinfo{year}{1971}).
	
	\bibitem[{\citenamefont{De~Gennes}(2018)}]{degennes-sc}
	\bibinfo{author}{\bibfnamefont{P.~G.} \bibnamefont{De~Gennes}},
	\emph{\bibinfo{title}{Superconductivity of Metals and Alloys.}}, Advanced
	Books Classics (\bibinfo{publisher}{Chapman and Hall/CRC},
	\bibinfo{address}{Boulder}, \bibinfo{year}{2018}), \bibinfo{edition}{2nd}
	ed., ISBN \bibinfo{isbn}{9780429965586}.
	
	\bibitem[{\citenamefont{Olshanii}(1998)}]{Olshani}
	\bibinfo{author}{\bibfnamefont{M.}~\bibnamefont{Olshanii}},
	\bibinfo{journal}{Phys. Rev. Lett.} \textbf{\bibinfo{volume}{81}},
	\bibinfo{pages}{938} (\bibinfo{year}{1998}),
	\urlprefix\url{https://link.aps.org/doi/10.1103/PhysRevLett.81.938}.
	
	\bibitem[{\citenamefont{Leggett}(2006)}]{Leggett}
	\bibinfo{author}{\bibfnamefont{A.~J.} \bibnamefont{Leggett}},
	\emph{\bibinfo{title}{Quantum Liquids : Bose condensation and Cooper pairing
			in condensed-matter systems}}, Oxford Graduate Texts
	(\bibinfo{publisher}{Oxford University Press}, \bibinfo{address}{Oxford, UK},
	\bibinfo{year}{2006}), ISBN \bibinfo{isbn}{0198526431}.
	
	\bibitem[{\citenamefont{Silvert}(1964)}]{PE-Silvert}
	\bibinfo{author}{\bibfnamefont{W.}~\bibnamefont{Silvert}},
	\bibinfo{journal}{Rev. Mod. Phys.} \textbf{\bibinfo{volume}{36}},
	\bibinfo{pages}{251} (\bibinfo{year}{1964}),
	\urlprefix\url{https://link.aps.org/doi/10.1103/RevModPhys.36.251}.
	
	\bibitem[{\citenamefont{Onofrio}(2016)}]{Onofrio_2016}
	\bibinfo{author}{\bibfnamefont{R.}~\bibnamefont{Onofrio}},
	\bibinfo{journal}{Phys.-Uspekhi.} \textbf{\bibinfo{volume}{59}},
	\bibinfo{pages}{1129} (\bibinfo{year}{2016}),
	\urlprefix\url{https://doi.org/10.3367/ufne.2016.07.037873}.
	
	\bibitem[{\citenamefont{Ferrier-Barbut
			et~al.}(2014)\citenamefont{Ferrier-Barbut, Delehaye, Laurent, Grier, Pierce,
			Rem, Chevy, and Salomon}}]{doi:10.1126/science.1255380}
	\bibinfo{author}{\bibfnamefont{I.}~\bibnamefont{Ferrier-Barbut}},
	\bibinfo{author}{\bibfnamefont{M.}~\bibnamefont{Delehaye}},
	\bibinfo{author}{\bibfnamefont{S.}~\bibnamefont{Laurent}},
	\bibinfo{author}{\bibfnamefont{A.~T.} \bibnamefont{Grier}},
	\bibinfo{author}{\bibfnamefont{M.}~\bibnamefont{Pierce}},
	\bibinfo{author}{\bibfnamefont{B.~S.} \bibnamefont{Rem}},
	\bibinfo{author}{\bibfnamefont{F.}~\bibnamefont{Chevy}}, \bibnamefont{and}
	\bibinfo{author}{\bibfnamefont{C.}~\bibnamefont{Salomon}},
	\bibinfo{journal}{Science} \textbf{\bibinfo{volume}{345}},
	\bibinfo{pages}{1035} (\bibinfo{year}{2014}),
	\eprint{https://www.science.org/doi/pdf/10.1126/science.1255380},
	\urlprefix\url{https://www.science.org/doi/abs/10.1126/science.1255380}.
	
	\bibitem[{\citenamefont{Kim and Chien}(2018)}]{Tom-18}
	\bibinfo{author}{\bibfnamefont{T.}~\bibnamefont{Kim}} \bibnamefont{and}
	\bibinfo{author}{\bibfnamefont{C.-C.} \bibnamefont{Chien}},
	\bibinfo{journal}{Phys. Rev. A} \textbf{\bibinfo{volume}{97}},
	\bibinfo{pages}{033628} (\bibinfo{year}{2018}),
	\urlprefix\url{https://link.aps.org/doi/10.1103/PhysRevA.97.033628}.
	
	\bibitem[{\citenamefont{Parajuli et~al.}(2023)\citenamefont{Parajuli, Pecak,
			and Chien}}]{BFpaper}
	\bibinfo{author}{\bibfnamefont{B.}~\bibnamefont{Parajuli}},
	\bibinfo{author}{\bibfnamefont{D.}~\bibnamefont{Pecak}}, \bibnamefont{and}
	\bibinfo{author}{\bibfnamefont{C.-C.} \bibnamefont{Chien}},
	\bibinfo{journal}{Phys. Rev. A} \textbf{\bibinfo{volume}{107}},
	\bibinfo{pages}{023308} (\bibinfo{year}{2023}),
	\urlprefix\url{https://link.aps.org/doi/10.1103/PhysRevA.107.023308}.
	
	\bibitem[{\citenamefont{Chin et~al.}(2010)\citenamefont{Chin, Grimm, Julienne,
			and Tiesinga}}]{RevModPhys.82.1225}
	\bibinfo{author}{\bibfnamefont{C.}~\bibnamefont{Chin}},
	\bibinfo{author}{\bibfnamefont{R.}~\bibnamefont{Grimm}},
	\bibinfo{author}{\bibfnamefont{P.}~\bibnamefont{Julienne}}, \bibnamefont{and}
	\bibinfo{author}{\bibfnamefont{E.}~\bibnamefont{Tiesinga}},
	\bibinfo{journal}{Rev. Mod. Phys.} \textbf{\bibinfo{volume}{82}},
	\bibinfo{pages}{1225} (\bibinfo{year}{2010}),
	\urlprefix\url{https://link.aps.org/doi/10.1103/RevModPhys.82.1225}.
	
	\bibitem[{\citenamefont{Fedichev et~al.}(1996)\citenamefont{Fedichev, Kagan,
			Shlyapnikov, and Walraven}}]{theory-optical-feshbach1}
	\bibinfo{author}{\bibfnamefont{P.~O.} \bibnamefont{Fedichev}},
	\bibinfo{author}{\bibfnamefont{Y.}~\bibnamefont{Kagan}},
	\bibinfo{author}{\bibfnamefont{G.~V.} \bibnamefont{Shlyapnikov}},
	\bibnamefont{and} \bibinfo{author}{\bibfnamefont{J.~T.~M.}
		\bibnamefont{Walraven}}, \bibinfo{journal}{Phys. Rev. Lett.}
	\textbf{\bibinfo{volume}{77}}, \bibinfo{pages}{2913} (\bibinfo{year}{1996}),
	\urlprefix\url{https://link.aps.org/doi/10.1103/PhysRevLett.77.2913}.
	
	\bibitem[{\citenamefont{Bauer et~al.}(2009)\citenamefont{Bauer, Lettner, Vo,
			Rempe, and Dürr}}]{exp-opt-mag-fesbach}
	\bibinfo{author}{\bibfnamefont{D.~M.} \bibnamefont{Bauer}},
	\bibinfo{author}{\bibfnamefont{M.}~\bibnamefont{Lettner}},
	\bibinfo{author}{\bibfnamefont{C.}~\bibnamefont{Vo}},
	\bibinfo{author}{\bibfnamefont{G.}~\bibnamefont{Rempe}}, \bibnamefont{and}
	\bibinfo{author}{\bibfnamefont{S.}~\bibnamefont{Dürr}},
	\bibinfo{journal}{Nat. Phys.} \textbf{\bibinfo{volume}{5}},
	\bibinfo{pages}{339} (\bibinfo{year}{2009}),
	\urlprefix\url{https://doi.org/10.1038}.
	
	\bibitem[{\citenamefont{Arunkumar et~al.}(2019)\citenamefont{Arunkumar,
			Jagannathan, and Thomas}}]{PhysRevLett.122.040405}
	\bibinfo{author}{\bibfnamefont{N.}~\bibnamefont{Arunkumar}},
	\bibinfo{author}{\bibfnamefont{A.}~\bibnamefont{Jagannathan}},
	\bibnamefont{and} \bibinfo{author}{\bibfnamefont{J.~E.}
		\bibnamefont{Thomas}}, \bibinfo{journal}{Phys. Rev. Lett.}
	\textbf{\bibinfo{volume}{122}}, \bibinfo{pages}{040405}
	(\bibinfo{year}{2019}),
	\urlprefix\url{https://link.aps.org/doi/10.1103/PhysRevLett.122.040405}.
	
	\bibitem[{\citenamefont{Mukherjee et~al.}(2017)\citenamefont{Mukherjee, Yan,
			Patel, Hadzibabic, Yefsah, Struck, and Zwierlein}}]{PhysRevLett.118.123401}
	\bibinfo{author}{\bibfnamefont{B.}~\bibnamefont{Mukherjee}},
	\bibinfo{author}{\bibfnamefont{Z.}~\bibnamefont{Yan}},
	\bibinfo{author}{\bibfnamefont{P.~B.} \bibnamefont{Patel}},
	\bibinfo{author}{\bibfnamefont{Z.}~\bibnamefont{Hadzibabic}},
	\bibinfo{author}{\bibfnamefont{T.}~\bibnamefont{Yefsah}},
	\bibinfo{author}{\bibfnamefont{J.}~\bibnamefont{Struck}}, \bibnamefont{and}
	\bibinfo{author}{\bibfnamefont{M.~W.} \bibnamefont{Zwierlein}},
	\bibinfo{journal}{Phys. Rev. Lett.} \textbf{\bibinfo{volume}{118}},
	\bibinfo{pages}{123401} (\bibinfo{year}{2017}),
	\urlprefix\url{https://link.aps.org/doi/10.1103/PhysRevLett.118.123401}.
	
	\bibitem[{\citenamefont{Hueck et~al.}(2018)\citenamefont{Hueck, Luick, Sobirey,
			Siegl, Lompe, and Moritz}}]{PhysRevLett.120.060402}
	\bibinfo{author}{\bibfnamefont{K.}~\bibnamefont{Hueck}},
	\bibinfo{author}{\bibfnamefont{N.}~\bibnamefont{Luick}},
	\bibinfo{author}{\bibfnamefont{L.}~\bibnamefont{Sobirey}},
	\bibinfo{author}{\bibfnamefont{J.}~\bibnamefont{Siegl}},
	\bibinfo{author}{\bibfnamefont{T.}~\bibnamefont{Lompe}}, \bibnamefont{and}
	\bibinfo{author}{\bibfnamefont{H.}~\bibnamefont{Moritz}},
	\bibinfo{journal}{Phys. Rev. Lett.} \textbf{\bibinfo{volume}{120}},
	\bibinfo{pages}{060402} (\bibinfo{year}{2018}),
	\urlprefix\url{https://link.aps.org/doi/10.1103/PhysRevLett.120.060402}.
	
	\bibitem[{\citenamefont{Mitra et~al.}(2018)\citenamefont{Mitra, Brown,
			Guardado-Sanchez, Kondov, Devakul, Huse, Schauss, and
			Bakr}}]{MitraDebayan2018Qgmo}
	\bibinfo{author}{\bibfnamefont{D.}~\bibnamefont{Mitra}},
	\bibinfo{author}{\bibfnamefont{P.~T.} \bibnamefont{Brown}},
	\bibinfo{author}{\bibfnamefont{E.}~\bibnamefont{Guardado-Sanchez}},
	\bibinfo{author}{\bibfnamefont{S.~S.} \bibnamefont{Kondov}},
	\bibinfo{author}{\bibfnamefont{T.}~\bibnamefont{Devakul}},
	\bibinfo{author}{\bibfnamefont{D.~A.} \bibnamefont{Huse}},
	\bibinfo{author}{\bibfnamefont{P.}~\bibnamefont{Schauss}}, \bibnamefont{and}
	\bibinfo{author}{\bibfnamefont{W.~S.} \bibnamefont{Bakr}},
	\bibinfo{journal}{Nat. Phys.} \textbf{\bibinfo{volume}{14}},
	\bibinfo{pages}{173} (\bibinfo{year}{2018}), ISSN \bibinfo{issn}{1745-2473}.
	
	\bibitem[{\citenamefont{Koepsell et~al.}(2020)\citenamefont{Koepsell, Hirthe,
			Bourgund, Sompet, Vijayan, Salomon, Gross, and
			Bloch}}]{PhysRevLett.125.010403}
	\bibinfo{author}{\bibfnamefont{J.}~\bibnamefont{Koepsell}},
	\bibinfo{author}{\bibfnamefont{S.}~\bibnamefont{Hirthe}},
	\bibinfo{author}{\bibfnamefont{D.}~\bibnamefont{Bourgund}},
	\bibinfo{author}{\bibfnamefont{P.}~\bibnamefont{Sompet}},
	\bibinfo{author}{\bibfnamefont{J.}~\bibnamefont{Vijayan}},
	\bibinfo{author}{\bibfnamefont{G.}~\bibnamefont{Salomon}},
	\bibinfo{author}{\bibfnamefont{C.}~\bibnamefont{Gross}}, \bibnamefont{and}
	\bibinfo{author}{\bibfnamefont{I.}~\bibnamefont{Bloch}},
	\bibinfo{journal}{Phys. Rev. Lett.} \textbf{\bibinfo{volume}{125}},
	\bibinfo{pages}{010403} (\bibinfo{year}{2020}),
	\urlprefix\url{https://link.aps.org/doi/10.1103/PhysRevLett.125.010403}.
	
	\bibitem[{\citenamefont{Hartke et~al.}(2020)\citenamefont{Hartke, Oreg, Jia,
			and Zwierlein}}]{PhysRevLett.125.113601}
	\bibinfo{author}{\bibfnamefont{T.}~\bibnamefont{Hartke}},
	\bibinfo{author}{\bibfnamefont{B.}~\bibnamefont{Oreg}},
	\bibinfo{author}{\bibfnamefont{N.}~\bibnamefont{Jia}}, \bibnamefont{and}
	\bibinfo{author}{\bibfnamefont{M.}~\bibnamefont{Zwierlein}},
	\bibinfo{journal}{Phys. Rev. Lett.} \textbf{\bibinfo{volume}{125}},
	\bibinfo{pages}{113601} (\bibinfo{year}{2020}),
	\urlprefix\url{https://link.aps.org/doi/10.1103/PhysRevLett.125.113601}.
	
	\bibitem[{\citenamefont{Hartke et~al.}(2022)\citenamefont{Hartke, Oreg,
			Turnbaugh, Jia, and Zwierlein}}]{hartke2022direct}
	\bibinfo{author}{\bibfnamefont{T.}~\bibnamefont{Hartke}},
	\bibinfo{author}{\bibfnamefont{B.}~\bibnamefont{Oreg}},
	\bibinfo{author}{\bibfnamefont{C.}~\bibnamefont{Turnbaugh}},
	\bibinfo{author}{\bibfnamefont{N.}~\bibnamefont{Jia}}, \bibnamefont{and}
	\bibinfo{author}{\bibfnamefont{M.}~\bibnamefont{Zwierlein}},
	\bibinfo{journal}{arXiv preprint arXiv:2208.05948}  (\bibinfo{year}{2022}).
	
	\bibitem[{\citenamefont{Gupta et~al.}(2003)\citenamefont{Gupta, Hadzibabic,
			Zwierlein, Stan, Dieckmann, Schunck, van Kempen, Verhaar, and
			Ketterle}}]{zweirlein-2003}
	\bibinfo{author}{\bibfnamefont{S.}~\bibnamefont{Gupta}},
	\bibinfo{author}{\bibfnamefont{Z.}~\bibnamefont{Hadzibabic}},
	\bibinfo{author}{\bibfnamefont{M.~W.} \bibnamefont{Zwierlein}},
	\bibinfo{author}{\bibfnamefont{C.~A.} \bibnamefont{Stan}},
	\bibinfo{author}{\bibfnamefont{K.}~\bibnamefont{Dieckmann}},
	\bibinfo{author}{\bibfnamefont{C.~H.} \bibnamefont{Schunck}},
	\bibinfo{author}{\bibfnamefont{E.~G.~M.} \bibnamefont{van Kempen}},
	\bibinfo{author}{\bibfnamefont{B.~J.} \bibnamefont{Verhaar}},
	\bibnamefont{and} \bibinfo{author}{\bibfnamefont{W.}~\bibnamefont{Ketterle}},
	\bibinfo{journal}{Science} \textbf{\bibinfo{volume}{300}},
	\bibinfo{pages}{1723} (\bibinfo{year}{2003}),
	\eprint{https://www.science.org/doi/pdf/10.1126/science.1085335},
	\urlprefix\url{https://www.science.org/doi/abs/10.1126/science.1085335}.
	
	\bibitem[{\citenamefont{Schunck et~al.}(2007)\citenamefont{Schunck, Shin,
			Schirotzek, Zwierlein, and Ketterle}}]{zweirlein-2007}
	\bibinfo{author}{\bibfnamefont{C.~H.} \bibnamefont{Schunck}},
	\bibinfo{author}{\bibfnamefont{Y.}~\bibnamefont{Shin}},
	\bibinfo{author}{\bibfnamefont{A.}~\bibnamefont{Schirotzek}},
	\bibinfo{author}{\bibfnamefont{M.~W.} \bibnamefont{Zwierlein}},
	\bibnamefont{and} \bibinfo{author}{\bibfnamefont{W.}~\bibnamefont{Ketterle}},
	\bibinfo{journal}{Science} \textbf{\bibinfo{volume}{316}},
	\bibinfo{pages}{867} (\bibinfo{year}{2007}),
	\eprint{https://www.science.org/doi/pdf/10.1126/science.1140749},
	\urlprefix\url{https://www.science.org/doi/abs/10.1126/science.1140749}.
	
	\bibitem[{\citenamefont{Mukherjee et~al.}(2019)\citenamefont{Mukherjee, Patel,
			Yan, Fletcher, Struck, and Zwierlein}}]{zweirlein-2019-1}
	\bibinfo{author}{\bibfnamefont{B.}~\bibnamefont{Mukherjee}},
	\bibinfo{author}{\bibfnamefont{P.~B.} \bibnamefont{Patel}},
	\bibinfo{author}{\bibfnamefont{Z.}~\bibnamefont{Yan}},
	\bibinfo{author}{\bibfnamefont{R.~J.} \bibnamefont{Fletcher}},
	\bibinfo{author}{\bibfnamefont{J.}~\bibnamefont{Struck}}, \bibnamefont{and}
	\bibinfo{author}{\bibfnamefont{M.~W.} \bibnamefont{Zwierlein}},
	\bibinfo{journal}{Phys. Rev. Lett.} \textbf{\bibinfo{volume}{122}},
	\bibinfo{pages}{203402} (\bibinfo{year}{2019}),
	\urlprefix\url{https://link.aps.org/doi/10.1103/PhysRevLett.122.203402}.
	
	\bibitem[{\citenamefont{Yan et~al.}(2019)\citenamefont{Yan, Patel, Mukherjee,
			Fletcher, Struck, and Zwierlein}}]{zweirlein-2019-2}
	\bibinfo{author}{\bibfnamefont{Z.}~\bibnamefont{Yan}},
	\bibinfo{author}{\bibfnamefont{P.~B.} \bibnamefont{Patel}},
	\bibinfo{author}{\bibfnamefont{B.}~\bibnamefont{Mukherjee}},
	\bibinfo{author}{\bibfnamefont{R.~J.} \bibnamefont{Fletcher}},
	\bibinfo{author}{\bibfnamefont{J.}~\bibnamefont{Struck}}, \bibnamefont{and}
	\bibinfo{author}{\bibfnamefont{M.~W.} \bibnamefont{Zwierlein}},
	\bibinfo{journal}{Phys. Rev. Lett.} \textbf{\bibinfo{volume}{122}},
	\bibinfo{pages}{093401} (\bibinfo{year}{2019}),
	\urlprefix\url{https://link.aps.org/doi/10.1103/PhysRevLett.122.093401}.
	
\end{thebibliography}

\end{document}